\documentclass[letter]{aa}

\usepackage{txfonts}
\usepackage{graphicx}
\usepackage{natbib}
\usepackage{aalongtable}
\usepackage{url}
\bibpunct[; ]{(}{)}{,}{a}{}{;} 
\usepackage{subcaption}
\usepackage{appendix}
\usepackage[normalem]{ulem}

\newenvironment{absolutelynopagebreak}
  {\par\nobreak\vfil\penalty0\vfilneg
   \vtop\bgroup}
  {\par\xdef\tpd{\the\prevdepth}\egroup
   \prevdepth=\tpd}

\begin{document}

\title{Mid-infrared evolution of $\eta$~Car from 1968 to 2018\thanks{Based on observations collected at ESO's Very Large Telescope under Prog-IDs: 074.A-9016(A), 0101.D-0077(A). Based on observations made with ESO Telescopes at the La Silla Paranal Observatory under Prog-IDs: 60.A-9126(A,C,E,I), 69.D-0304(B),71.D-0049(A).}
\fnmsep 
%\thanks{This work was co-funded under the Marie Curie Actions of the European Commission (FP7-COFUND).}\\ 
} 
\titlerunning{Mid-infrared evolution of $\eta$~Car from 1968 to 2018}
%\subtitle{Spectroscopic and photometric} 

\author{
A.\ Mehner\inst{1}
\and W.-J.\ de Wit\inst{1}
\and D.\ Asmus\inst{2}
\and P.W.\ Morris\inst{3}
\and C.\ Agliozzo\inst{4} 
\and M.J.\ Barlow\inst{5}
\and T.R.\ Gull\inst{6}
\and D.J.\ Hillier\inst{7}
\and G.\ Weigelt\inst{8}
} 
%\offprints{A. Mehner, \email{amehner@eso.org}}

\institute{ESO -- European Organisation for Astronomical Research in the Southern Hemisphere, Alonso de Cordova 3107, Vitacura, Santiago de Chile, Chile 
  \and Department of Physics \& Astronomy, University of Southampton, Southampton, SO17 1BJ, UK
  \and California Institute of Technology, IPAC, M/C 100-22, Pasadena, CA 91125, USA
  \and ESO -- European Organisation for Astronomical Research in the Southern Hemisphere, Karl-Schwarzschild-Stra{\ss}e 2,  85748 Garching, Germany
  \and Department of Physics \& Astronomy, University College London, Gower Street, London WC1E 6BT, UK
  \and NASA Goddard Space Flight Center, Code 667, Greenbelt, MD 20771, USA
  \and Department of Physics \& Astronomy, University of Pittsburgh, 3941 O'Hara Street, Pittsburgh, PA 15260, USA
  \and Max Planck Institute for Radio Astronomy, Auf dem H\"{u}gel 69, 53121, Bonn, Germany
}   

%\date{Received 2 November 1992 / Accepted 7 January 1993}

\abstract {Eta~Car is one of the most luminous and massive stars in our Galaxy and is the brightest mid-infrared (mid-IR) source in the sky, outside our solar system.  Since the late 1990s the central source has dramatically brightened at ultraviolet  and optical wavelengths. This might be explained by a decrease in circumstellar dust extinction. We aim to establish the mid-IR flux evolution and further our understanding of the star's ultraviolet and optical brightening. Mid-IR images from $8-20~\mu$m were obtained in 2018 with VISIR at the Very Large Telescope.  Archival data from 2003 and 2005 are retrieved from the ESO Science Archive Facility and historical records are collected from publications. We present the highest angular resolution mid-IR images of $\eta$~Car to date at the corresponding wavelengths ($\geq 0.22\arcsec$). We reconstruct the mid-IR evolution of the spectral energy distribution of the spatially integrated Homunculus nebula from 1968 to 2018 and find no long-term changes. Eta~Car's bolometric luminosity has been stable over the past five decades. We do not observe a long-term decrease in the mid-IR flux densities that could be associated with the brightening at ultraviolet and optical wavelengths, but circumstellar dust must be declining in our line-of-sight only. Short-term flux variations within about 25\% of the mean levels could be present.} 


\maketitle

\section{Introduction}
\label{intro}

\begin{figure}
\centering
\resizebox{1\hsize}{!}
{\includegraphics[width=1.06\textwidth]{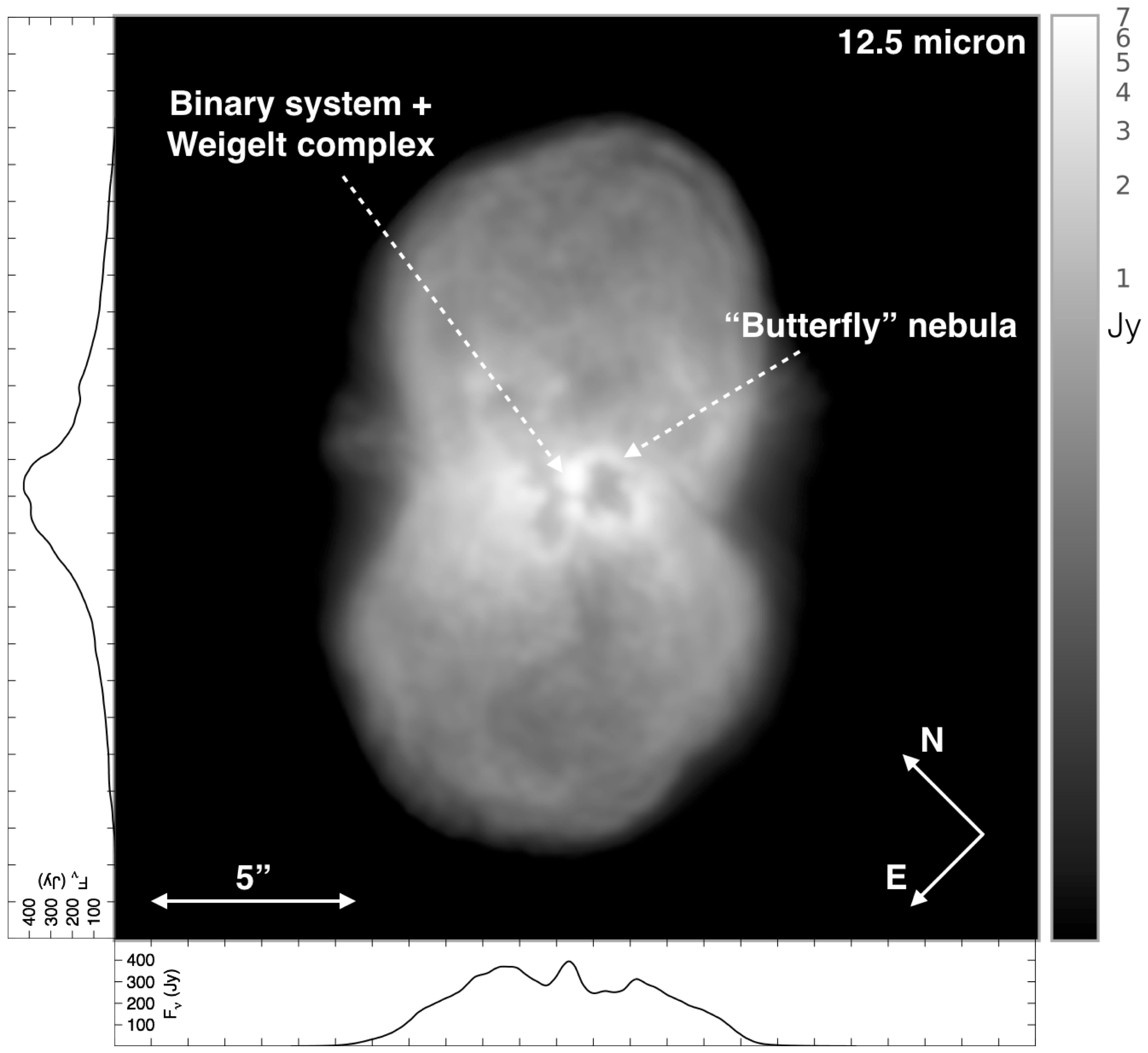}  } 
\caption{VISIR image of the Homunculus nebula at $12.5~\mu$m in 2018, tracing the thermal emission from heated dust and the \ion{H}{I} 7-6 emission. The field-of-view is $25\arcsec \times 25\arcsec$, the spatial resolution is $0.3\arcsec$. The flux density in Jy per detector pixel ($0.045\arcsec$/pixel) is shown as well as the integrated flux along the image axes. The brightest knot in the center of the nebula is due to a shell surrounding the star, an inner torus or disk, or a pin-wheel structure, and includes the Weigelt complex \citep{1986A&A...163L...5W}. The two bright loops form the ``Butterfly'' nebula, where 50\% of the integrated flux originates from. 
}
     \label{figure:B12.4}
\end{figure}

Eta~Car is the archetype of unstable massive stars, eruptive mass loss, supernova (SN) impostors, and a reference for understanding the precursor eruptions that lead to superluminous SNe. 
The star has a current mass of about $100~M_{\odot}$ and is in an eccentric binary system (\citealt{1997NewA....2..107D,2012ASSL..384.....D} and references therein).
During an eruptive event in the 19th century (see \citealt{2004JAD....10....6F} for the historical lightcurve), the system ejected $\geq 45~M_{\odot}$ of material (\citealt{2017ApJ...842...79M}; see also \citealt{2003AJ....125.1458S}), forming the bipolar Homunculus nebula. 

Eta~Car is often described as an extreme case of a Luminous Blue Variable (LBV). LBVs are evolved massive stars with initial masses $\geq20~M_{\odot}$ that exhibit instabilities that are not understood, resulting in enhanced mass loss (\citealt{1984IAUS..105..233C,1997ASPC..120..387C,1994PASP..106.1025H,1997ASPC..120.....N}). 
%They experience outbursts with enhanced mass loss during which they make transitions in the Hertzsprung-Russell diagram from their quiescent hot state ($T_{\textnormal{\scriptsize{eff}}}\sim16\,000-30\,000$~K) to lower temperatures ($T_{\textnormal{\scriptsize{eff}}}\sim8\,000$~K).
%LBV outbursts with visual magnitude variations of $1-2$~mag are commonly referred to as classical LBV outbursts. During giant eruptions, such as $\eta$~Car in the 1840s (e.g., \citealt{1997ARA&A..35....1D,2012ASSL..384.....D}) and P\,Cygni in the 17th century (e.g., \citealt{1988IrAJ...18..163D,1992A&A...257..153L}), the visual magnitude increases by more than 2~mag. 
They have been considered to be stars in transition to the Wolf-Rayet stage (e.g., \citealt{1983A&A...120..113M,1994A&A...290..819L}), but more recent observational and theoretical work suggests that some LBVs could be the immediate progenitors of SNe (e.g., \citealt{2006A&A...460L...5K,2008A&A...483L..47T,2009Natur.458..865G,2007ApJ...666.1116S,2013A&A...550L...7G}).
%The most promising explanations for the LBV instability mechanism involve radiation pressure instabilities, but also turbulent pressure instabilities, vibrations and dynamical instabilities, and binarity cannot be ruled out (see \citealt{1994PASP..106.1025H} for an overview). The high stellar luminosities near the Eddington limit probably enable instabilities  \citep{1993MNRAS.263..375G,2001MNRAS.326..126S,2002A&A...393..543V,2006ApJ...645L..45S,2012A&A...538A..40G,2014arXiv1402.0257G,2015ASSL..412..113O}. 

A crucial parameter for understanding $\eta$~Car is the luminosity since this provides a mass estimate via the Eddington limit. 
Most of $\eta$~Car's visible and ultraviolet (UV) light is absorbed by circumstellar dust and re-radiated in the infrared (IR; \citealt{1971MNRAS.154..415D,1972NPhS..236...46D,1969ApL.....4..221P,1969ApJ...156L..45W,1973MNRAS.161..281R,1974ApJ...190L..69S}). The star's luminosity can thus be obtained from the IR spectral energy distribution (SED).
Eta~Car is unparalleled as an IR radiation source and its brightness makes observations with sensitive IR satellite instrumentation impossible. 
Most of the $0.25-0.44~M_{\odot}$ of cool dust is located in the central $5\arcsec$ region in a ``disrupted'' equatorial structure \citep{1999Natur.402..502M,2017ApJ...842...79M}, also referred to as the ``Butterfly'' nebula \citep{2005A&A...435.1043C}. 
Early estimates of $\eta$~Car's total luminosity are on the order of $L = 5 \times 10^6~L_{\odot}$ \citep{1968ApJ...152L..89N,1969ApJ...156L..45W,1969ApL.....4..221P,1973MNRAS.161..281R,1974ApJ...190L..69S,1995A&A...297..168C,1997ARA&A..35....1D}.
\citet{2017ApJ...842...79M} conclude that $\eta$~Car's total luminosity in the 1970s was $L = 4.1 \times 10^6~L_{\odot}$, for the commonly adopted distance of 2.3~kpc. A distance revision to 2.6~kpc could imply a $25-30\%$ increase in luminosity \citep{2018RNAAS...2c.133D}.  
 The luminosity of the secondary star is a factor of 10 lower  \citep{2010ApJ...710..729M}.

Some authors have suggested a change in the mid-IR flux since the 1970s. \citet{1987ApJ...321..937R} reported a 30\% decrease in the $10.5~\mu$m silicate emission compared to values for the 1970s.
\citet{1995MNRAS.273..354S} found a significant decrease in mid-IR flux in the 1990s compared to the 1970s. \citet{2017ApJ...842...79M} derive a total luminosity of $L = 3.0 \times 10^6~L_{\odot}$ for the 1990s, a decline of 25\% in 20~yr. They argue that the decline in mid-IR flux indicates a reduction in circumstellar extinction either by the expansion of dust ``shells'' or dust destruction. This could cause the simultaneous brightening at UV and optical wavelengths  \citep{1999AJ....118.1777D,2005AJ....129..900D,2006AJ....132.2717M}.

We present newly obtained and previously unpublished mid-IR observations, which are the basis of a re-evaluation of $\eta$~Car's mid-IR evolution from 1968 to 2018. 
In Section \ref{obs} we describe the new mid-IR images. We present our findings and re-analyze $\eta$~Car's  SED over a period of 50 years in Section \ref{results_discussion}. In Section \ref{conclusion} we conclude that $\eta$~Car has not undergone any appreciable long-term luminosity changes during this period.

\section{Mid-IR observations in 2005 and 2018 with VISIR and in 2003 with TIMMI2}
\label{obs}

\begin{table}
\caption{Mid-IR flux densities of $\eta$~Car's spatially integrated Homunculus nebula (2003--2018).\label{table:mid-infrared}}
\begin{tabular}{lcccc}
\hline\hline
Filter & Wavelength & Date & Flux & Error flux \\ % table heading
 &  &  & density & density$^a$ \\ % table heading

        &  ($\mu$m) & & (Jy) & (Jy) \\
        \hline
& & VISIR & & \\
J7.9      &       7.78            &   2018-05-21    &       9648	&	30      \\
PAH1            &       8.59     &    2018-04-28   &      19280	&	40             \\
ArIII &	8.99		&	2018-12-10	&	31140	&	80	\\
SIV\_1 &	9.82		&	2018-05-21	&	51030	&	220	\\
SIV &       10.49   &    2018-04-29   &       54520	&	160 \\
PAH2    &       11.25   &   2018-12-10    &       66660 		&	210      \\
B12.4	&	 12.47	&	2018-04-29	&	53930	&		230 \\
NeII    &       12.81   &    2018-12-10   &       63120	&	470   \\
NeII\_2   &       13.04   &    2018-12-10   &       58410	&	450   \\
Q1      &       17.65	&    2018-05-11   &	 93040	&	2690      \\
Q2      &       18.72	&    2018-05-11   &	100810	&	3550     \\
Q3      &       19.50	&    2018-05-11   &	97380	&	9940   \\
\hline
PAH2    &       11.25   &   2005-01-23    &       61350	&	390     \\
NeII$^{b}$    &       12.81   &    2005-01-23   &       44420	& 1050  \\
\hline
& &  TIMMI2$^{c}$ & & \\
M	&	4.6     &   2003    &   3590	&	1230    \\
N7.9	&	7.9    &   2003    &   13260	&	3880  \\
N8.9	&	8.7    &   2003    &   23650	&	7810  \\
N9.8	&	9.6     &   2003    &   40590	&	2840 \\
N10.4	&	10.3     &   2003    &   47710	&	5060  \\
N11.9	&	11.6    &   2003    &   63360	&	6980  \\
N12.9	&	12.3     &   2003    &   60000	&	6100  \\
$[$NeII$]$	&	12.8    &   2003    &  80960	&	8170 \\
Q1	&	17.8     &   2003    &   122190	&	15190  \\
\hline
\multicolumn{5}{l}{$^a$ Formal errors from the flux calibration. Systematic  }\\
\multicolumn{5}{l}{uncertainties due to sky variability and detector artifacts can}\\
\multicolumn{5}{l}{be larger than $10\%$ for $N$ band and $20\%$ for $Q$ band.} \\
\multicolumn{5}{l}{$^b$ Observations are strongly impacted by ghosts.} \\
\multicolumn{5}{l}{$^c$ Average of observations in January, March, and May 2003.} \\
\multicolumn{5}{l}{The data are badly affected by stripe patterns.} \\\end{tabular}
\end{table}

\begin{figure*}
\centering
\resizebox{\hsize}{!}{\includegraphics{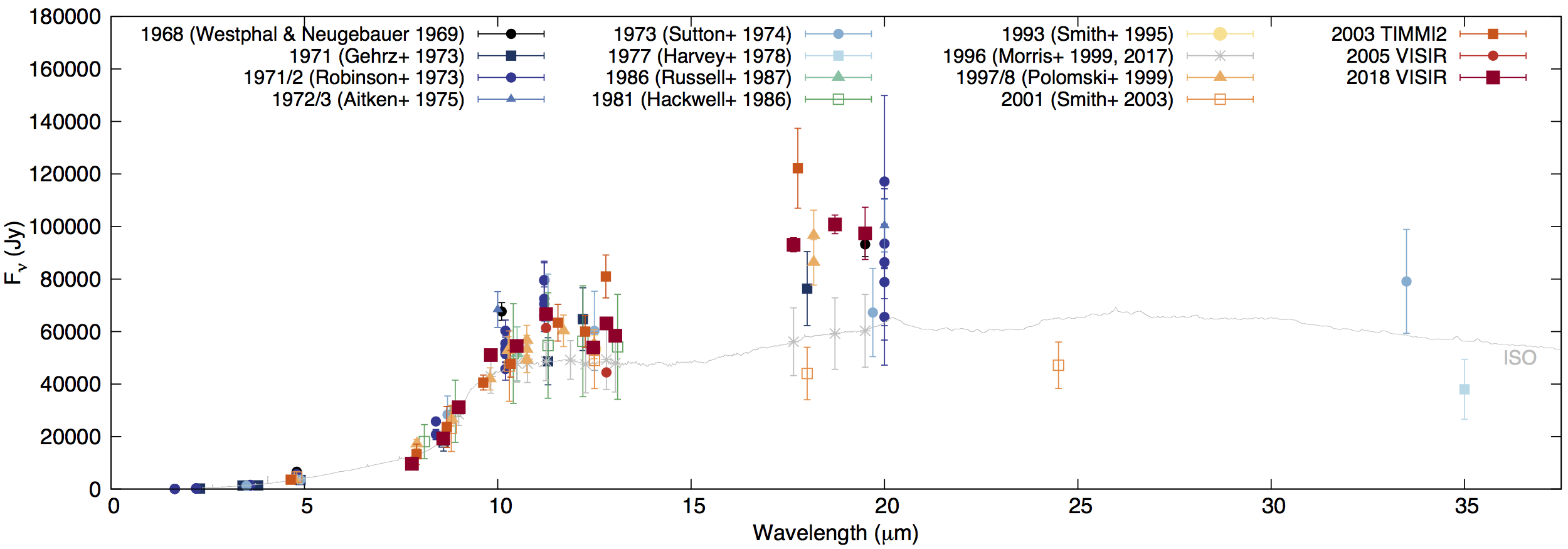}}
     \caption{Mid-IR photometry of the integrated Homunculus nebula from 1968 to 2018. Colored symbols represent our new and historic mid-IR measurements performed in the 1960s \citep{1969ApJ...156L..45W}, 1970s \citep{1973ApL....13...89G,1973MNRAS.161..281R,1975MNRAS.172..141A,1974ApJ...190L..69S,1978A&A....70..165H},  1980s \citep{1986ApJ...311..380H,1987ApJ...321..937R}, 1990s \citep{1995MNRAS.273..354S,1999AJ....118.2369P,1999Natur.402..502M,2017ApJ...842...79M}, and 2000s \citep{2003AJ....125.1458S}. 
Calibration uncertainties may be larger than reported  due to unaccounted systematic uncertainties, sky variability, detector artifacts, and partial saturation. Open symbols represent values estimated from isophotal contour maps \citep{1986ApJ...311..380H,2003AJ....125.1458S}.
    %Sloan: 2003ApJS..147..379S. The flux difference between the two versions of the displayed ISO spectrum originates from assumptions within the data reduction process. Note that the spectrum from \citealt{2003ApJS..147..379S} shows artifacts, e.g, the bump at $15~\mu$m.
%\citet{1999Natur.402..502M} note that the values published by \citet{1978A&A....70..165H} are too low due to calibration issues.
}
     \label{figure:SED}
\end{figure*}

Eta~Car was observed in 2018 with the upgraded VLT imager and spectrometer for the mid-IR (VISIR; \citealt{2004Msngr.117...12L,2015Msngr.159...15K}).
The VISIR AQUARIUS detector provides a field of view of $38.0\arcsec \times 38.0\arcsec$ with a pixel scale of $0.045\arcsec$. 
We sampled the mid-IR wavelength range with several filters from $7.8-19.5~\mu$m, see Table \ref{table:mid-infrared}, Table \ref{table:journal_visir}, and Figure \ref{figure: VISIR images}. The observations were carried out at an airmass below 1.3 and at a precipitable water vapor between $1.2-3.5$~mm, ensuring good image quality.
The burst read-out mode was used, which results in a large number of short exposure images with instantaneous point spread functions, corresponding to the momentary atmospheric turbulence. High spatial resolution images can be obtained by re-centering and adding the individual short-exposure images.
We reach spatial resolutions close to the diffraction limits in the given filters, namely $0.25-0.30\arcsec$ in $N$ band and $\sim 0.50\arcsec$ in $Q$ band.
For the flux calibration, the science observations were either preceded or followed by a mid-IR standard star  (HD89682; \citealt{1999AJ....117.1864C}) obtained with the same setup as the science observations.

The data reduction was performed with a custom-made \textsc{python} pipeline.
The photometry was determined using classical aperture photometry with an aperture radius of $11\arcsec$, which encompasses the entire Homunculus nebula.
VISIR absolute flux calibration is accurate to 10\% in the $N$ band and 20\% in the $Q$ band. The dominating source of uncertainty is the mid-IR sky variability. 
The count levels in the central bright core of the Homunculus nebula (1\arcsec\ in diameter, see Figure \ref{figure:B12.4}) are outside the linear regime for most filters, which implies that we may be underestimating the total flux by a few percent. We do not attempt to correct for this, because longward of $8~\mu$m less than 10\% and longward of $10~\mu$m less than 5\% of the flux is produced by this central core.  
The resulting mid-infrared spectral flux densities and statistical uncertainties are listed in Table \ref{table:mid-infrared}. 

In what follows, we reconstruct the mid-IR SED with our 2018 observations and literature data (references are provided in the caption of Figure \ref{figure:SED}). In order to be as complete as possible, we also make use of mid-IR images in the ESO Science Archive Facility with unpublished photometry.\footnote{\url{http://archive.eso.org}}
Eta~Car was observed with VISIR in 2005 in the filters PAH1, PAH2, NeII, and Q2 with the former Boeing detector. Standard star observations were only obtained in the filters PAH2 and NeII, see Table \ref{table:journal_visir}. The pixel scale is 0.075\arcsec\ and the images have spatial resolutions of $\sim 0.35\arcsec$. Detector artifacts, such as strong striping and ghosts, result in poor flux measurements. 
In January, March, and May 2003, $\eta$~Car was observed with the Thermal infrared Multi-Mode instrument (TIMMI2, \citealt{2000SPIE.4008.1132R}) at the La Silla 3.6-m Telescope, see Table \ref{table:journal_timmi2}. The TIMMI2 array shows smear and stripe patterns for bright sources, which lead to large errors in the flux calibration. We have averaged the flux values of all TIMMI2 observations and present the standard deviation as errors in Table \ref{table:mid-infrared}. 
%The error in $Q$ band is much larger than in $N$ band because of low signal-to-noise standard star observations, varying sky background, and atmospheric transmission. 
The TIMMI2 data should be interpreted with caution.

\begin{figure*}
\centering
\resizebox{\hsize}{!}{\includegraphics{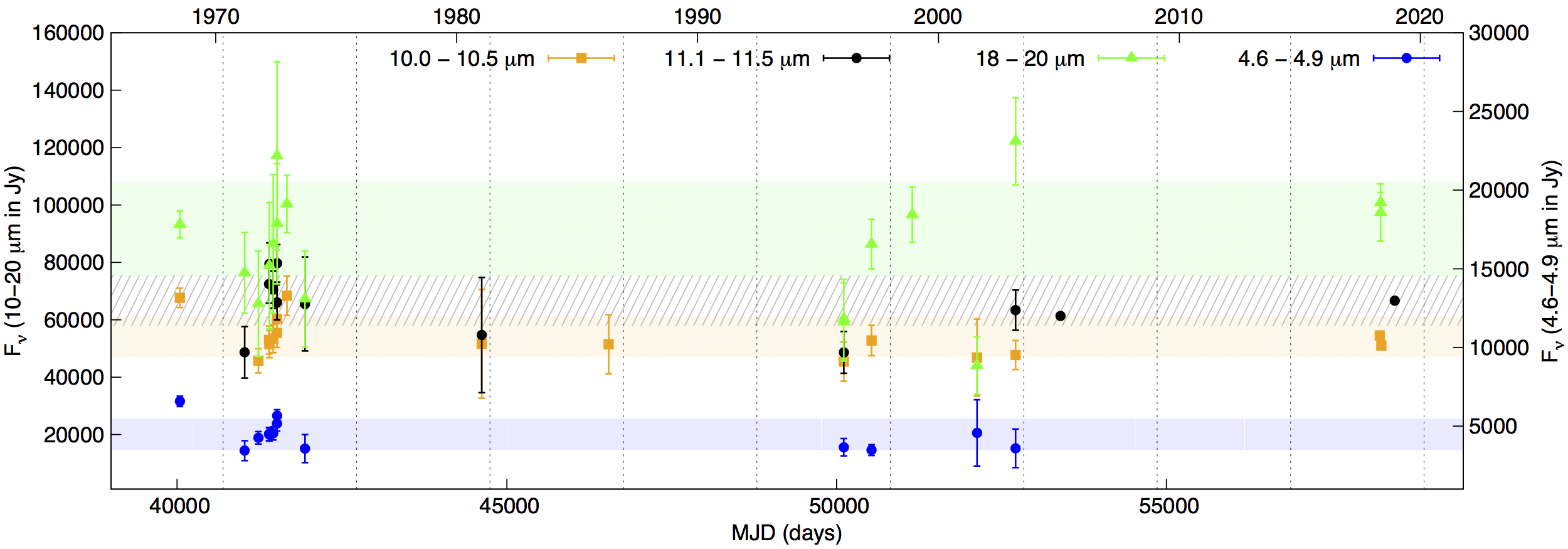}}
     \caption{Time evolution of the mid-IR flux of $\eta$~Car's Homunculus nebula from 1968 to 2018 in four wavelength regions, chosen for best temporal coverage. Vertical dashed lines indicate periastron passages. The $1\sigma$ region of the average flux for each wavelength region is shown (averages exclude the lower ISO flux values). There is no evidence for a long-term change, but variations with the orbital period cannot be ruled out.
}
     \label{figure:flux_versus_time}
\end{figure*}

\section{Results}
\label{results_discussion}

\subsection{Eta~Car's flux evolution in the mid-IR}
\label{results:SED}

Figure \ref{figure:SED} compares the 2018 mid-IR photometry from our VISIR images with previous observations in the period 1968--2005. Figure \ref{figure:flux_versus_time} displays the time evolution of four mid-IR wavelength regions over 50~years and their averages. With a few exceptions, individual values are consistent with each other within $1\sigma$ ($22\%$ for $M$ band, $13\%$ for $N$ band, and $19\%$ for $Q$ band) of the average. The figures demonstrate long-term stability of the mid-IR flux between 5 and 20~$\mu$m over 50~years within the uncertainty of mid-IR flux calibration. 

The timescales for the condensation and destruction of dust grains depend on the chemistry and shielding conditions, which are complex in $\eta$~Car's Homunculus nebula. Ionizing UV and X-ray radiation escaping from the central binary vary with orbital phase, but we cannot confirm periodic changes in the mid-IR because the observations were obtained at a mix of orbital phases, the orbital coverage is sparse, and the calibration uncertainties are large. The observations in 1971/72 indicate a trend to higher flux values in mid-cycle, similar to what is observed in the radio \citep{2005ASPC..332..126W}, but \citet{1973MNRAS.161..281R} state that there is no evidence for variations from October 1971 to July 1972 larger than the measurement uncertainties. \citet{1999AJ....118.2369P} find that the $N$ band and $18~\mu$m flux density increase between March 1997 and November 1998, i.e., around periastron. 
The emission of the central bright core ($1\arcsec$ in diameter) contributes to less than 10\% at $8~\mu$m, 5\% at $10~\mu$m, 4\% at $13~\mu$m, and 2\% at $19~\mu$m. Any variations due to recent dust formation or destruction close to the star is hidden in the uncertainties of the nebula-encompassing photometry. 

\citet{2017ApJ...842...79M} reported that $\eta$~Car's mid-IR flux  decreased by 25\% over the last decades.\footnote{\citet{2017ApJ...842...79M} display too high values in their figure 7 for the $11~\mu$m photometry by \citet{1973ApL....13...89G} due to a transcription error. \citet{1973ApL....13...89G} contains an inconsistency of a factor of about two between the flux values at $11~\mu$m reported in their table 1 and figure 1. We use the values from their table 1.}  Their result was based on data obtained in 1996 with the spectrometers onboard the Infrared Space Observatory (ISO; \citealt{1996A&A...315L..27K}). %Spectroscopic observations with the Short Wavelength Spectrometer (SWS01; \citealt{1996A&A...315L..49D}) were obtained from $2.4-45.2~\mu$m with a spectral resolution ranging from 1000 to 2000. 
While the absolute spectro-photometric calibration of the ISO spectra was hampered by different detector materials with their own signal-dependent susceptibilities to non-linearities and memory effects (e.g., \citealt{2004A&A...414..677V,2017ApJ...842...79M}),
%; ISO handbook\footnote{\url{https://www.cosmos.esa.int/web/iso/iso-handbook}})
the reported decline may be at least partly due to intrinsic orbital variations. 
A detailed color-temperature analysis with orbital phase will reveal the thermal characteristics of  short-term variations, but such a study will require a more homogeneous set of spatial and temporal monitoring observations.

%It appears that the treatment of the memory effects could have resulted in too low fluxes at $16-28~\mu$m (by about $25\%$, cf.\ \citealt{1999AJ....118.2369P}).
%We also include part of the ISO Long Wavelength Spectrometer (LWS) spectra. 
%The authors provide an uncertainty of 23\% in the  $12-26.5~\mu$m range, 30\% in the $26.5-29.0~\mu$m range, and 25\% in the $29.0-45.2~\mu$m range. 
%Flux calibration uncertainties are about 30\%. 

In contrast to \citet{1987ApJ...321..937R,1995MNRAS.273..354S} and \citet{2017ApJ...842...79M},  our SED re-construction and investigation of the time evolution does not reveal a noticeable long-term decrease in the mid-IR flux from $8-20~\mu$m since the first available mid-IR photometry in the late 1960s (see Figures \ref{figure:SED} and \ref{figure:flux_versus_time}).
The data support a rather stable mid-IR flux over 50~years, within the large uncertainties of ground-based mid-IR photometry and excluding variations within $\sim25\%$ of the mean levels. As a consequence, the UV and optical brightening since the late 1990s does not correlate with the integrated mid-IR flux. This is an important finding as the lack of any systematic mid-IR flux change rules out large changes in $\eta$~Car's luminosity.  We derive a luminosity of about  $4.6 \times 10^6~L_\odot$ for $\eta$~Car, using the 2.3~$\mu$m, 3.4~$\mu$m, and 4.9~$\mu$m photometry of \citet{1973ApL....13...89G}, the 2005 and 2018 VISIR mid-IR photometry, and the flux values at $35-175~\mu$m by \citet{1978A&A....70..165H} and at 450~$\mu$m and 850~$\mu$m by \citet{2010MNRAS.401L..48G}.

\subsection{On dust extinction}

Ever since the 1940s, $\eta$~Car has brightened steadily at optical wavelengths \citep{2004JAD....10....6F} and an accelerated brightening has been observed since the late 1990s \citep{1999AJ....118.1777D}. Today, $\eta$~Car at UV and visual wavelengths is more than 2~mag brighter than in the 1990s and the change is remarkably gray \citep{2018ApJ...858..109D,2019MNRAS.484.1325D}.
Because $\eta$~Car is close to the Eddington Limit, in principle this cannot constitute an increase in bolometric luminosity. 
\citet{1999AJ....118.1777D,2005AJ....129..900D} and \citet{2006AJ....132.2717M} attribute this brightening to a rapid decrease of the circumstellar extinction. The material around $\eta$~Car is not spherically distributed. The central region is seen through a dust condensation, which strongly attenuates the central source \citep{1992A&A...262..153H,1995RMxAC...2...11W}. \citet{1999AJ....118.1777D} argue that movement of a dusty condensation that intercepts our line-of-sight is unlikely, leaving dust destruction or a decrease of the dust formation rate close to the star as possible explanations. 
A possible cause may also be small changes in the stellar parameters, which alter the shape of the wind-wind shock cone such that we are no longer looking through the newly formed dust.
\citet{2019MNRAS.484.1325D} propose the dissipation of a dusty clump (``coronagraph'') in our line-of-sight.

The variability in the near-IR from 1972--2013 was studied by \citet{2004MNRAS.352..447W} and \citet{2014A&A...564A..14M}. The star shows a long-term brightening in $JHK$, whereas the $L$-band emission remains basically constant. 
The $4-8~\mu$m flux also shows no long-term variation (\citealt{1987ApJ...321..937R}; Figure \ref{figure:flux_versus_time}). Given that longward of $4~\mu$m thermal radiation dominates the SED rather than extinction, one would expect a significant decrease in flux in the $10-20~\mu$m range only in case of a net destruction of warm ($\sim 200$~K) dust grains. Nonetheless, Figure \ref{figure:flux_versus_time} indicates a short-term increase in emission levels during mid-cycle in the 1970s and during periastron in 1998.

The extinction of the UV and optical light is probably caused predominantly by dust within the inner core.
In the VISIR J7.9 filter image, the central core is resolved in at least three knots (Figure \ref{figure: VISIR images}). In the PAH2 and NeII filters, for which we have comparison VISIR images in 2005, this core is unresolved. For an aperture of 0.5\arcsec\ radius around the central bright source, we derive $F_{\nu,~ \mathrm{PAH2,~2005}} = 3130\pm50$~Jy and  $F_{\nu,~ \mathrm{NeII,~2005}} = 2710\pm180$~Jy, compared with  $F_{\nu,~\mathrm{PAH2,~2018}} = 3140\pm10$~Jy and  $F_{\nu,~\mathrm{NeII,~2018}} = 3200\pm30$~Jy. Additional systematic errors are about 10\%. There appears to be no detectable change in mid-IR flux of the inner core in the last 13~yr. This implies either an equilibrium or very weak dust formation and destruction close to the star. 
If the geometry of this central knot would simply expand, then the optical depth $\tau(\mathrm{UV, optical})$ would decline, depending on grain properties (chemistry, shape, size). In this case, and if $\tau(\mathrm{IR}) < 1$, we would expect a small effect ($<$~2--10\% depending on wavelength and hidden in the uncertainties) in the integrated mid-IR flux since the IR emission is determined by the larger nebula. Any stochastic increase or decrease in line-of-sight dust changes the UV and optical extinction instantly, but not the IR emission as the dust particles are heated and will radiate.

%At longer wavelengths, \citet{2010MNRAS.401L..48G} studied the submillimeter variability at $870~\mu$m and attributed the variations to the ionized stellar wind rather than dust emission. \citet{2003MNRAS.338..425D} and \citet{2005ASPC..332..126W} monitored the radio emission and find a cycle corresponding to the binary period. They interpret the emission to arise in the dense outflow  photoionized by the companion star.

\subsection{Resolved dusty circumstellar structures}
\label{geometry}

Several authors have suggested a disk-, ring-, or torus-like structure of $5-6\arcsec$ in diameter around the central star, which absorbs, scatters, and re-radiates the stellar radiation in the IR
\citep{1973ApL....13...89G,1974ApJ...190L..69S,1979ApJ...233..145H,1978MmSAI..49..589W,1995RMxAC...2...27R,1995MNRAS.273..354S,1999AJ....118.2369P}. \citet{1999Natur.402..502M} show that most of the dust in the equatorial region is located in toroidal structures, probably created in shock-heated gas of the 19th century eruption. Some dust may also be located in an inner unresolved torus or disk and/or pinwheel-like structures created by colliding winds in the orbital plane  \citep{2016A&A...594A.106W}.
\citet{2002ApJ...567L..77S} present mid-IR images from $4.8-24.5~\mu$m which support the presence of a circumstellar torus or disk.
\citet{2011AJ....141..202A} argue that this ``Butterfly'' nebula \citep{2005A&A...435.1043C} is not a coherent physical structure or equatorial torus but spatially separate clumps and filaments ejected at different times.
Radial velocity information from ALMA observations support the general picture that the loops are a pinched torus in the orbital plane, perpendicular to the Homunculus lobes  \citep{2018MNRAS.474.4988S}. The direction of the pinched material matches periastron/apastron orientation \citep{2013MNRAS.436.3820M} and the companion may have played a role in disrupting the torus soon after its ejection. 

The geometry of the dust surrounding $\eta$~Car is quite complex, see Figure \ref{figure:B12.4} (also \citealt{2002ApJ...567L..77S}). In the center, two bright ring-like structures form the ``Butterfly'' nebula. Most of the dust is located in this central region (50\% of the flux originates within a 3\arcsec\ radius from the central source, 80\% within a 5\arcsec\ radius). In the mid-IR, one can identify strong density or temperature gradients and knots along these loops, but also spatially coherent structures.  
%There are no PAHs detected in $\eta$~Car's spatially unresolved spectrum and the cavities inside these loops are thus unlikely to be photo-dissociation regions.
%The lack of PAHs does not exclude PDRs, this is chemistry dependent.

Comparison between the 2005 and 2018 VISIR images show that the two prominent loops are expanding at a rate of up to $0.01\arcsec~$yr$^{-1}$ projected on sky. This corresponds to projected velocities of $\sim 100$~km~s$^{-1}$ at a distance of 2.3~kpc, consistent with velocities in the equatorial plane \citep{2001AJ....121.1569D}. 
%The presence of moving ejecta close to the star, make line-of-sight extinction variability over the time span of decades plausible (cf.\ \citealt{1999AJ....118.1777D}). 
The rings seem to expand without losing their overall appearance, in support of physically coherent structures. 
The furthest extent of the ``Butterfly'' nebula is about 2\arcsec\ from the central source, which points toward an ejection during or in the decades before the 19th century eruption. 
We do not find any brightness changes of the inner loop-like structures between observations separated by 13~years. For an aperture of 3\arcsec\ radius, we derive $F_{\nu,~\mathrm{PAH2,~2005}} = 34170\pm520$~Jy and  $F_{\nu,~\mathrm{NeII,~2005}} = 30100\pm2010$~Jy, compared with  $F_{\nu,~ \mathrm{PAH2,~2018}} = 34500\pm150$~Jy and  $F_{\nu,~ \mathrm{NeII,~2018}} = 32020\pm280$~Jy.
The response time for the dust formation or destruction at these distances from the star is not clear. However, the constant flux and overall appearance of these loops may imply that there is no recent dust formation or destruction and the material is simply expanding, as these structures are likely optically thin. 

Clues to the history and origin of $\eta$~Car's mass ejection phases can be found in its circumstellar material. Hydrodynamical simulations show that a spherical mass ejection into a massive pre-existing torus of gas and dust could result in the present-day bi-polar geometry of the Homunculus nebula \citep{1995ApJ...441L..77F,1998AJ....116..829D}. The massive torus may have been created shortly before the 19th century eruption, e.g., through non-conservative mass transfer, leaving an unstable core that then erupted.

\section{Conclusions}
\label{conclusion}

The mid-IR flux densities of $\eta$~Car's Homunculus nebula from $8-20~\mu$m show no long-term decline since the first available mid-IR photometry in 1968. Eta~Car's luminosity has thus probably been stable over the last five decades ($\sim4.6 \times 10^6~L_\odot$, adopting a distance of 2.3~kpc).
Mid-IR observations were obtained irregularly and at different orbital phases. The large uncertainties of the mid-IR photometry (10\% in $N$ band and 20\% in $Q$ band) prohibits the confirmation of smaller short-term fluctuations or variations with the orbital period. 

Contrary to previous publications, we find no long-term decline of mid-IR photometric levels, which would have indicated a reduction in circumstellar extinction and could have explained the increase in UV and optical brightness. 
The most likely scenario to explain $\eta$~Car's brightening at UV and optical wavelengths, and its stability in the mid-IR, is that the extinction caused by circumstellar dust is declining in our line-of-sight only.

\begin{acknowledgements} 
We thank the anonymous referee for the constructive feedback. MJB acknowledges support from ERC grant SNDUST 694520. We have used SketchAndCalc to calculate the areas in isophotal contour maps (E.\ M.\ Dobbs; \url{www.SketchAndCalc.com}).
%Based on observations with ISO, an ESA project with instruments funded by ESA Member States (especially the PI countries: France, Germany, the Netherlands and the United Kingdom) and with the participation of ISAS and NASA. 
\end{acknowledgements}

%\bibliographystyle{aa}
%\bibliography{ms}

\begin{thebibliography}{65}
\expandafter\ifx\csname natexlab\endcsname\relax\def\natexlab#1{#1}\fi

\bibitem[{{Aitken} \& {Jones}(1975)}]{1975MNRAS.172..141A}
{Aitken}, D.~K. \& {Jones}, B. 1975, \mnras, 172, 141

\bibitem[{{Artigau} {et~al.}(2011){Artigau}, {Martin}, {Humphreys}, {Davidson},
  {Chesneau}, \& {Smith}}]{2011AJ....141..202A}
{Artigau}, {\'E}., {Martin}, J.~C., {Humphreys}, R.~M., {et~al.} 2011, \aj,
  141, 202

\bibitem[{{Chesneau} {et~al.}(2005){Chesneau}, {Min}, {Herbst}, {Waters},
  {Hillier}, {Leinert}, {de Koter}, {Pascucci}, {Jaffe}, {K{\"o}hler},
  {Alvarez}, {van Boekel}, {Brandner}, {Graser}, {Lagrange}, {Lenzen}, {Morel},
  \& {Sch{\"o}ller}}]{2005A&A...435.1043C}
{Chesneau}, O., {Min}, M., {Herbst}, T., {et~al.} 2005, \aap, 435, 1043

\bibitem[{{Cohen} {et~al.}(1999){Cohen}, {Walker}, {Carter}, {Hammersley},
  {Kidger}, \& {Noguchi}}]{1999AJ....117.1864C}
{Cohen}, M., {Walker}, R.~G., {Carter}, B., {et~al.} 1999, \aj, 117, 1864

\bibitem[{{Conti}(1984)}]{1984IAUS..105..233C}
{Conti}, P.~S. 1984, in IAU Symposium, Vol. 105, Observational Tests of the
  Stellar Evolution Theory, ed. A.~{Maeder} \& A.~{Renzini}, 233

\bibitem[{{Conti}(1997)}]{1997ASPC..120..387C}
{Conti}, P.~S. 1997, in PASPC, Vol. 120, Luminous Blue Variables: Massive Stars
  in Transition, ed. A.~{Nota} \& H.~{Lamers}, 387

\bibitem[{{Cox} {et~al.}(1995){Cox}, {Mezger}, {Sievers}, {Najarro},
  {Bronfman}, {Kreysa}, \& {Haslam}}]{1995A&A...297..168C}
{Cox}, P., {Mezger}, P.~G., {Sievers}, A., {et~al.} 1995, \aap, 297, 168

\bibitem[{{Damineli} {et~al.}(1997){Damineli}, {Conti}, \&
  {Lopes}}]{1997NewA....2..107D}
{Damineli}, A., {Conti}, P.~S., \& {Lopes}, D.~F. 1997, New Astronomy, 2, 107

\bibitem[{{Damineli} {et~al.}(2019){Damineli}, {Fern{\'a}ndez-Laj{\'u}s},
  {Almeida}, {Corcoran}, {Damineli}, {Gull}, {Hamaguchi}, {Hillier},
  {Jablonski}, {Madura}, {Moffat}, {Navarete}, {Richardson}, {Ruiz}, {Salerno},
  {Scalia}, \& {Weigelt}}]{2019MNRAS.484.1325D}
{Damineli}, A., {Fern{\'a}ndez-Laj{\'u}s}, E., {Almeida}, L.~A., {et~al.} 2019,
  \mnras, 484, 1325

\bibitem[{{Davidson}(1971)}]{1971MNRAS.154..415D}
{Davidson}, K. 1971, \mnras, 154, 415

\bibitem[{{Davidson} {et~al.}(1999){Davidson}, {Gull}, {Humphreys},
  {Ishibashi}, {Whitelock}, {Berdnikov}, {McGregor}, {Metcalfe}, {Polomski}, \&
  {Hamuy}}]{1999AJ....118.1777D}
{Davidson}, K., {Gull}, T.~R., {Humphreys}, R.~M., {et~al.} 1999, \aj, 118,
  1777

\bibitem[{{Davidson} {et~al.}(2018{\natexlab{a}}){Davidson}, {Helmel}, \&
  {Humphreys}}]{2018RNAAS...2c.133D}
{Davidson}, K., {Helmel}, G., \& {Humphreys}, R.~M. 2018{\natexlab{a}}, RNAAS,
  2, 133

\bibitem[{{Davidson} \& {Humphreys}(1997)}]{1997ARA&A..35....1D}
{Davidson}, K. \& {Humphreys}, R.~M. 1997, \araa, 35, 1

\bibitem[{{Davidson} \& {Humphreys}(2012)}]{2012ASSL..384.....D}
{Davidson}, K. \& {Humphreys}, R.~M., eds. 2012, ASSL, Vol. 384, {Eta Carinae
  and the Supernova Impostors}

\bibitem[{{Davidson} {et~al.}(2018{\natexlab{b}}){Davidson}, {Ishibashi},
  {Martin}, \& {Humphreys}}]{2018ApJ...858..109D}
{Davidson}, K., {Ishibashi}, K., {Martin}, J.~C., \& {Humphreys}, R.~M.
  2018{\natexlab{b}}, \apj, 858, 109

\bibitem[{{Davidson} {et~al.}(2005){Davidson}, {Martin}, {Humphreys},
  {Ishibashi}, {Gull}, {Stahl}, {Weis}, {Hillier}, {Damineli}, {Corcoran}, \&
  {Hamann}}]{2005AJ....129..900D}
{Davidson}, K., {Martin}, J., {Humphreys}, R.~M., {et~al.} 2005, \aj, 129, 900

\bibitem[{{Davidson} \& {Ostriker}(1972)}]{1972NPhS..236...46D}
{Davidson}, K. \& {Ostriker}, J.~P. 1972, Nature Physical Science, 236, 46

\bibitem[{{Davidson} {et~al.}(2001){Davidson}, {Smith}, {Gull}, {Ishibashi}, \&
  {Hillier}}]{2001AJ....121.1569D}
{Davidson}, K., {Smith}, N., {Gull}, T.~R., {Ishibashi}, K., \& {Hillier},
  D.~J. 2001, \aj, 121, 1569

\bibitem[{{Dwarkadas} \& {Balick}(1998)}]{1998AJ....116..829D}
{Dwarkadas}, V.~V. \& {Balick}, B. 1998, \aj, 116, 829

\bibitem[{{Frank} {et~al.}(1995){Frank}, {Balick}, \&
  {Davidson}}]{1995ApJ...441L..77F}
{Frank}, A., {Balick}, B., \& {Davidson}, K. 1995, \apjl, 441, L77

\bibitem[{{Frew}(2004)}]{2004JAD....10....6F}
{Frew}, D.~J. 2004, Journal of Astronomical Data, 10

\bibitem[{{Gal-Yam} \& {Leonard}(2009)}]{2009Natur.458..865G}
{Gal-Yam}, A. \& {Leonard}, D.~C. 2009, \nat, 458, 865

\bibitem[{{Gehrz} {et~al.}(1973){Gehrz}, {Ney}, {Becklin}, \&
  {Neugebauer}}]{1973ApL....13...89G}
{Gehrz}, R.~D., {Ney}, E.~P., {Becklin}, E.~E., \& {Neugebauer}, G. 1973,
  \aplett, 13, 89

\bibitem[{{Gomez} {et~al.}(2010){Gomez}, {Vlahakis}, {Stretch}, {Dunne},
  {Eales}, {Beelen}, {Gomez}, \& {Edmunds}}]{2010MNRAS.401L..48G}
{Gomez}, H.~L., {Vlahakis}, C., {Stretch}, C.~M., {et~al.} 2010, \mnras, 401,
  L48

\bibitem[{{Groh} {et~al.}(2013){Groh}, {Meynet}, \&
  {Ekstr{\"o}m}}]{2013A&A...550L...7G}
{Groh}, J.~H., {Meynet}, G., \& {Ekstr{\"o}m}, S. 2013, \aap, 550, L7

\bibitem[{{Hackwell} {et~al.}(1986){Hackwell}, {Gehrz}, \&
  {Grasdalen}}]{1986ApJ...311..380H}
{Hackwell}, J.~A., {Gehrz}, R.~D., \& {Grasdalen}, G.~L. 1986, \apj, 311, 380

\bibitem[{{Harvey} {et~al.}(1978){Harvey}, {Hoffmann}, \&
  {Campbell}}]{1978A&A....70..165H}
{Harvey}, P.~M., {Hoffmann}, W.~F., \& {Campbell}, M.~F. 1978, \aap, 70, 165

\bibitem[{{Hillier} \& {Allen}(1992)}]{1992A&A...262..153H}
{Hillier}, D.~J. \& {Allen}, D.~A. 1992, \aap, 262, 153

\bibitem[{{Humphreys} \& {Davidson}(1994)}]{1994PASP..106.1025H}
{Humphreys}, R.~M. \& {Davidson}, K. 1994, \pasp, 106, 1025

\bibitem[{{Hyland} {et~al.}(1979){Hyland}, {Robinson}, {Mitchell}, {Thomas}, \&
  {Becklin}}]{1979ApJ...233..145H}
{Hyland}, A.~R., {Robinson}, G., {Mitchell}, R.~M., {Thomas}, J.~A., \&
  {Becklin}, E.~E. 1979, \apj, 233, 145

\bibitem[{{K{\"a}ufl} {et~al.}(2015){K{\"a}ufl}, {Kerber}, {Asmus}, {Baksai},
  {Di Lieto}, {Duhoux}, {Heikamp}, {Hummel}, {Ives}, {Jakob}, {Kirchbauer},
  {Mehrgan}, {Momany}, {Pantin}, {Pozna}, {Riquelme}, {Sandrock},
  {Siebenmorgen}, {Smette}, {Stegmeier}, {Taylor}, {Tristram}, {Valdes}, {van
  den Ancker}, {Weilenmann}, \& {Wolff}}]{2015Msngr.159...15K}
{K{\"a}ufl}, H.~U., {Kerber}, F., {Asmus}, D., {et~al.} 2015, The Messenger,
  159, 15

\bibitem[{{Kessler} {et~al.}(1996){Kessler}, {Steinz}, {Anderegg}, {Clavel},
  {Drechsel}, {Estaria}, {Faelker}, {Riedinger}, {Robson}, {Taylor}, \&
  {Xim{\'e}nez de Ferr{\'a}n}}]{1996A&A...315L..27K}
{Kessler}, M.~F., {Steinz}, J.~A., {Anderegg}, M.~E., {et~al.} 1996, \aap, 315,
  L27

\bibitem[{{Kotak} \& {Vink}(2006)}]{2006A&A...460L...5K}
{Kotak}, R. \& {Vink}, J.~S. 2006, \aap, 460, L5

\bibitem[{{Lagage} {et~al.}(2004){Lagage}, {Pel}, {Authier}, {Belorgey},
  {Claret}, {Doucet}, {Dubreuil}, {Durand}, {Elswijk}, {Girardot}, {K{\"a}ufl},
  {Kroes}, {Lortholary}, {Lussignol}, {Marchesi}, {Pantin}, {Peletier},
  {Pirard}, {Pragt}, {Rio}, {Schoenmaker}, {Siebenmorgen}, {Silber}, {Smette},
  {Sterzik}, \& {Veyssiere}}]{2004Msngr.117...12L}
{Lagage}, P.~O., {Pel}, J.~W., {Authier}, M., {et~al.} 2004, The Messenger,
  117, 12

\bibitem[{{Langer} {et~al.}(1994){Langer}, {Hamann}, {Lennon}, {Najarro},
  {Pauldrach}, \& {Puls}}]{1994A&A...290..819L}
{Langer}, N., {Hamann}, W.-R., {Lennon}, M., {et~al.} 1994, \aap, 290, 819

\bibitem[{{Madura} {et~al.}(2013){Madura}, {Gull}, {Okazaki}, {Russell},
  {Owocki}, {Groh}, {Corcoran}, {Hamaguchi}, \&
  {Teodoro}}]{2013MNRAS.436.3820M}
{Madura}, T.~I., {Gull}, T.~R., {Okazaki}, A.~T., {et~al.} 2013, \mnras, 436,
  3820

\bibitem[{{Maeder}(1983)}]{1983A&A...120..113M}
{Maeder}, A. 1983, \aap, 120, 113

\bibitem[{{Martin} {et~al.}(2006){Martin}, {Davidson}, \&
  {Koppelman}}]{2006AJ....132.2717M}
{Martin}, J.~C., {Davidson}, K., \& {Koppelman}, M.~D. 2006, \aj, 132, 2717

\bibitem[{{Mehner} {et~al.}(2010){Mehner}, {Davidson}, {Ferland}, \&
  {Humphreys}}]{2010ApJ...710..729M}
{Mehner}, A., {Davidson}, K., {Ferland}, G.~J., \& {Humphreys}, R.~M. 2010,
  \apj, 710, 729

\bibitem[{{Mehner} {et~al.}(2014){Mehner}, {Ishibashi}, {Whitelock},
  {Nagayama}, {Feast}, {van Wyk}, \& {de Wit}}]{2014A&A...564A..14M}
{Mehner}, A., {Ishibashi}, K., {Whitelock}, P., {et~al.} 2014, \aap, 564, A14

\bibitem[{{Morris} {et~al.}(2017){Morris}, {Gull}, {Hillier}, {Barlow},
  {Royer}, {Nielsen}, {Black}, \& {Swinyard}}]{2017ApJ...842...79M}
{Morris}, P.~W., {Gull}, T.~R., {Hillier}, D.~J., {et~al.} 2017, \apj, 842, 79

\bibitem[{{Morris} {et~al.}(1999){Morris}, {Waters}, {Barlow}, {Lim}, {de
  Koter}, {Voors}, {Cox}, {de Graauw}, {Henning}, {Hony}, {Lamers}, {Mutschke},
  \& {Trams}}]{1999Natur.402..502M}
{Morris}, P.~W., {Waters}, L.~B.~F.~M., {Barlow}, M.~J., {et~al.} 1999, \nat,
  402, 502

\bibitem[{{Neugebauer} \& {Westphal}(1968)}]{1968ApJ...152L..89N}
{Neugebauer}, G. \& {Westphal}, J.~A. 1968, \apjl, 152, L89

\bibitem[{{Nota} \& {Lamers}(1997)}]{1997ASPC..120.....N}
{Nota}, A. \& {Lamers}, H., eds. 1997, ASPC, Vol. 120, {Luminous Blue
  Variables: Massive Stars in Transition}

\bibitem[{{Pagel}(1969)}]{1969ApL.....4..221P}
{Pagel}, B.~E.~T. 1969, \aplett, 4, 221

\bibitem[{{Polomski} {et~al.}(1999){Polomski}, {Telesco}, {Pi{\~n}a}, \&
  {Fisher}}]{1999AJ....118.2369P}
{Polomski}, E.~F., {Telesco}, C.~M., {Pi{\~n}a}, R.~K., \& {Fisher}, R.~S.
  1999, \aj, 118, 2369

\bibitem[{{Reimann} {et~al.}(2000){Reimann}, {Linz}, {Wagner}, {Relke},
  {Kaeufl}, {Dietzsch}, {Sperl}, \& {Hron}}]{2000SPIE.4008.1132R}
{Reimann}, H.-G., {Linz}, H., {Wagner}, R., {et~al.} 2000, in \procspie, Vol.
  4008, Optical and IR Telescope Instrumentation and Detectors, ed. M.~{Iye} \&
  A.~F. {Moorwood}, 1132--1143

\bibitem[{{Rigaut} \& {Gehring}(1995)}]{1995RMxAC...2...27R}
{Rigaut}, F. \& {Gehring}, G. 1995, in RMxAA, vol.~27, Vol.~2, RMxAA Conf.
  Ser., ed. V.~{Niemela}, N.~{Morrell}, \& A.~{Feinstein}, 27--35

\bibitem[{{Robinson} {et~al.}(1973){Robinson}, {Hyland}, \&
  {Thomas}}]{1973MNRAS.161..281R}
{Robinson}, G., {Hyland}, A.~R., \& {Thomas}, J.~A. 1973, \mnras, 161, 281

\bibitem[{{Russell} {et~al.}(1987){Russell}, {Lynch}, {Hackwell}, {Rudy},
  {Rossano}, \& {Castelaz}}]{1987ApJ...321..937R}
{Russell}, R.~W., {Lynch}, D.~K., {Hackwell}, J.~A., {et~al.} 1987, \apj, 321,
  937

\bibitem[{{Smith} {et~al.}(1995){Smith}, {Aitken}, {Moore}, {Roche}, {Puetter},
  \& {Pina}}]{1995MNRAS.273..354S}
{Smith}, C.~H., {Aitken}, D.~K., {Moore}, T.~J.~T., {et~al.} 1995, \mnras, 273,
  354

\bibitem[{{Smith} {et~al.}(2003){Smith}, {Gehrz}, {Hinz}, {Hoffmann}, {Hora},
  {Mamajek}, \& {Meyer}}]{2003AJ....125.1458S}
{Smith}, N., {Gehrz}, R.~D., {Hinz}, P.~M., {et~al.} 2003, \aj, 125, 1458

\bibitem[{{Smith} {et~al.}(2002){Smith}, {Gehrz}, {Hinz}, {Hoffmann},
  {Mamajek}, {Meyer}, \& {Hora}}]{2002ApJ...567L..77S}
{Smith}, N., {Gehrz}, R.~D., {Hinz}, P.~M., {et~al.} 2002, \apjl, 567, L77

\bibitem[{{Smith} {et~al.}(2018){Smith}, {Ginsburg}, \&
  {Bally}}]{2018MNRAS.474.4988S}
{Smith}, N., {Ginsburg}, A., \& {Bally}, J. 2018, \mnras, 474, 4988

\bibitem[{{Smith} {et~al.}(2007){Smith}, {Li}, {Foley}, {Wheeler}, {Pooley},
  {Chornock}, {Filippenko}, {Silverman}, {Quimby}, {Bloom}, \&
  {Hansen}}]{2007ApJ...666.1116S}
{Smith}, N., {Li}, W., {Foley}, R.~J., {et~al.} 2007, \apj, 666, 1116

\bibitem[{{Sutton} {et~al.}(1974){Sutton}, {Becklin}, \&
  {Neugebauer}}]{1974ApJ...190L..69S}
{Sutton}, E., {Becklin}, E.~E., \& {Neugebauer}, G. 1974, \apjl, 190, L69

\bibitem[{{Trundle} {et~al.}(2008){Trundle}, {Kotak}, {Vink}, \&
  {Meikle}}]{2008A&A...483L..47T}
{Trundle}, C., {Kotak}, R., {Vink}, J.~S., \& {Meikle}, W.~P.~S. 2008, \aap,
  483, L47

\bibitem[{{Van Malderen} {et~al.}(2004){Van Malderen}, {Decin}, {Kester},
  {Vandenbussche}, {Waelkens}, {Cami}, \& {Shipman}}]{2004A&A...414..677V}
{Van Malderen}, R., {Decin}, L., {Kester}, D., {et~al.} 2004, \aap, 414, 677

\bibitem[{{Warren-Smith} {et~al.}(1978){Warren-Smith}, {Scarrot}, {Murdin}, \&
  {Bingham}}]{1978MmSAI..49..589W}
{Warren-Smith}, R.~F., {Scarrot}, S.~M., {Murdin}, P., \& {Bingham}, R.~G.
  1978, Memorie della Societa Astronomica Italiana, 49, 589

\bibitem[{{Weigelt} {et~al.}(1995){Weigelt}, {Albrecht}, {Barbieri}, {Blades},
  {Boksenberg}, {Crane}, {Davidson}, {Deharveng}, {Disney}, {Jakobsen},
  {Kamperman}, {King}, {Macchetto}, {Mackay}, {Paresce}, {Baxter},
  {Greenfield}, {Jedrzejewski}, {Nota}, \& {Sparks}}]{1995RMxAC...2...11W}
{Weigelt}, G., {Albrecht}, R., {Barbieri}, C., {et~al.} 1995, in RMxAA,
  vol.~27, Vol.~2, RMxAA Conf. Ser., ed. V.~{Niemela}, N.~{Morrell}, \&
  A.~{Feinstein}, 11

\bibitem[{{Weigelt} \& {Ebersberger}(1986)}]{1986A&A...163L...5W}
{Weigelt}, G. \& {Ebersberger}, J. 1986, \aap, 163, L5

\bibitem[{{Weigelt} {et~al.}(2016){Weigelt}, {Hofmann}, {Schertl}, {Clementel},
  {Corcoran}, {Damineli}, {de Wit}, {Grellmann}, {Groh}, {Guieu}, {Gull},
  {Heininger}, {Hillier}, {Hummel}, {Kraus}, {Madura}, {Mehner}, {M{\'e}rand},
  {Millour}, {Moffat}, {Ohnaka}, {Patru}, {Petrov}, {Rengaswamy}, {Richardson},
  {Rivinius}, {Sch{\"o}ller}, {Teodoro}, \& {Wittkowski}}]{2016A&A...594A.106W}
{Weigelt}, G., {Hofmann}, K.-H., {Schertl}, D., {et~al.} 2016, \aap, 594, A106

\bibitem[{{Westphal} \& {Neugebauer}(1969)}]{1969ApJ...156L..45W}
{Westphal}, J.~A. \& {Neugebauer}, G. 1969, \apjl, 156, L45

\bibitem[{{White} {et~al.}(2005){White}, {Duncan}, {Chapman}, \&
  {Koribalski}}]{2005ASPC..332..126W}
{White}, S.~M., {Duncan}, R.~A., {Chapman}, J.~M., \& {Koribalski}, B. 2005, in
  ASP Conf. Ser., Vol. 332, The Fate of the Most Massive Stars, ed.
  R.~{Humphreys} \& K.~{Stanek}, 129

\bibitem[{{Whitelock} {et~al.}(2004){Whitelock}, {Feast}, {Marang}, \&
  {Breedt}}]{2004MNRAS.352..447W}
{Whitelock}, P.~A., {Feast}, M.~W., {Marang}, F., \& {Breedt}, E. 2004, \mnras,
  352, 447

\end{thebibliography}

\appendix
\onecolumn

\section{Mid-IR images with VISIR in 2018}
\label{appendix:images}

\begin{figure}[h]
\begin{minipage}{0.96\textwidth}
\begin{subfigure}[b]{1\textwidth}
\includegraphics[width=0.34\textwidth]{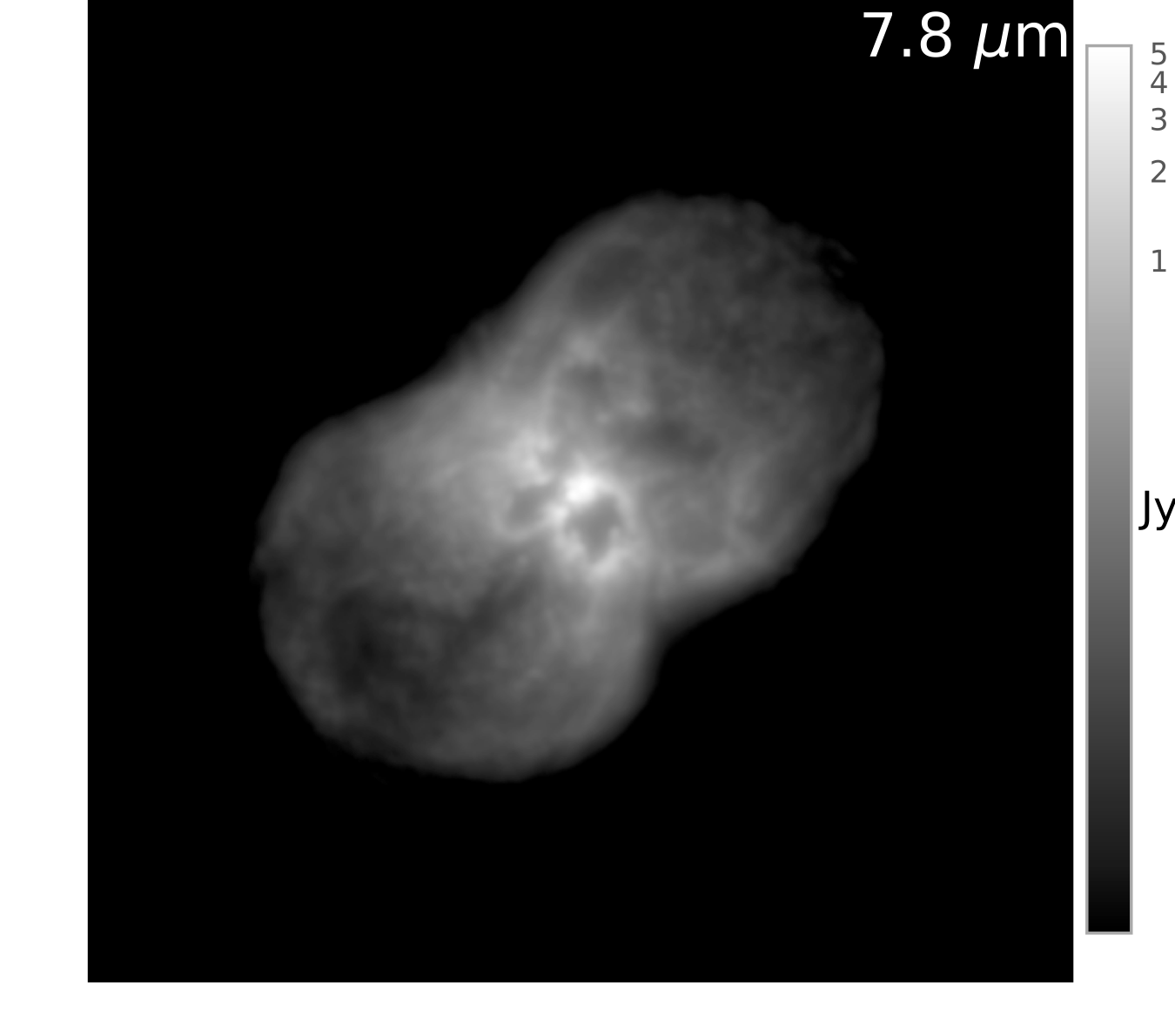}\hspace{0.0\textwidth}
\includegraphics[width=0.34\textwidth]{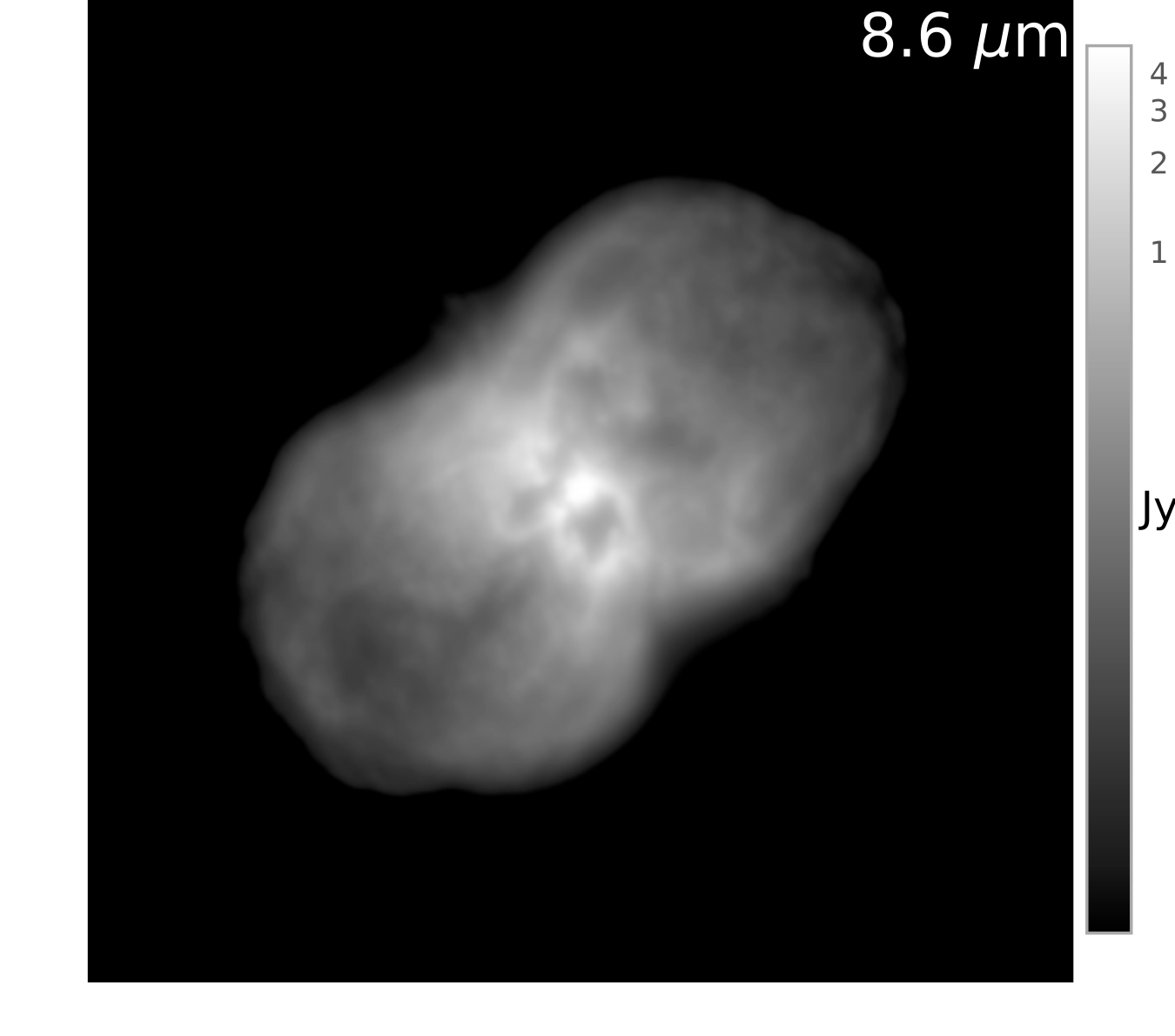}\hspace{0.0\textwidth}
\includegraphics[width=0.34\textwidth]{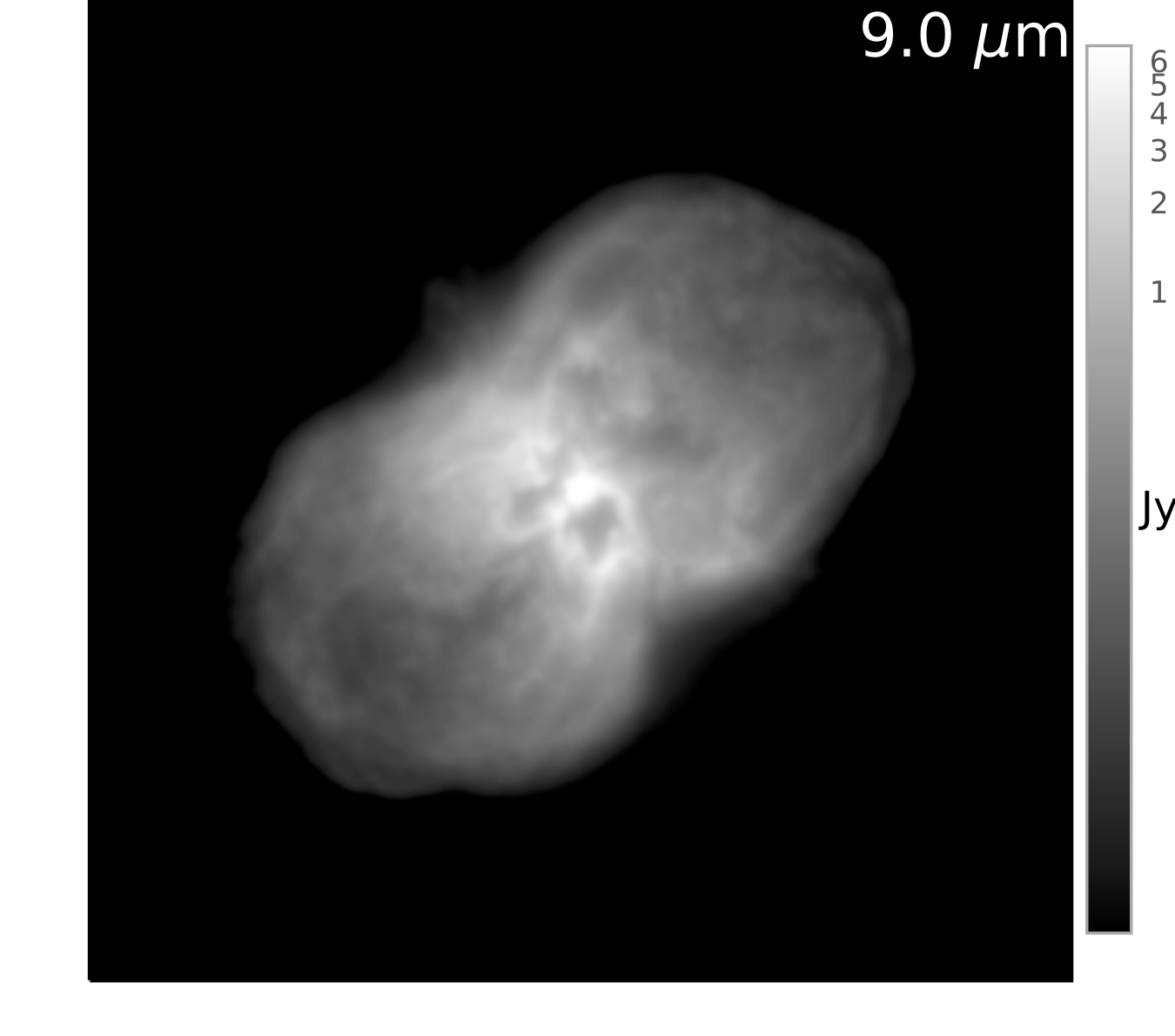}\vspace{0.0\textwidth}
\end{subfigure}
\end{minipage}
\begin{minipage}{0.96\textwidth}
\includegraphics[width=0.3425\textwidth]{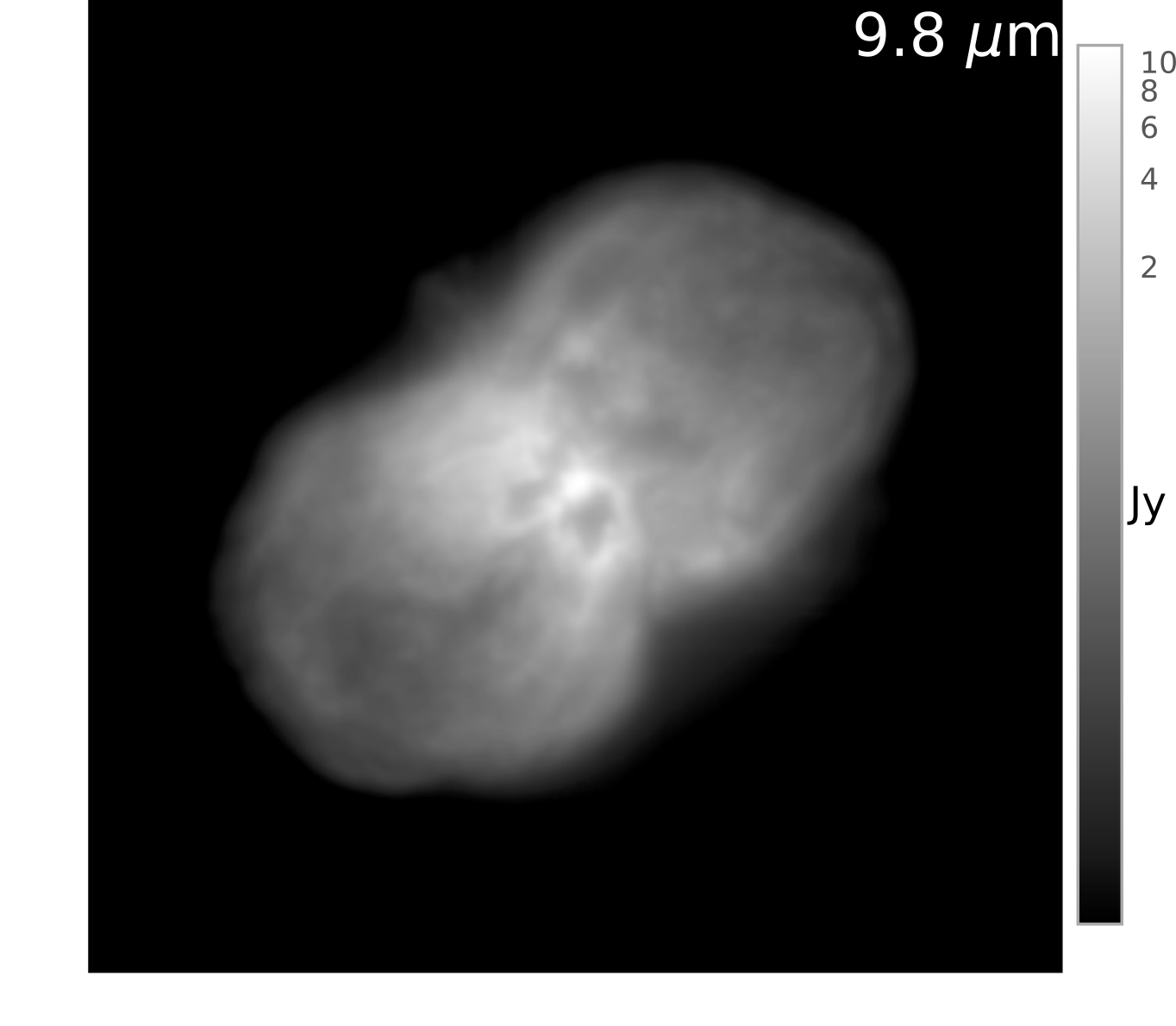}\hspace{0.0\textwidth}
\includegraphics[width=0.34\textwidth]{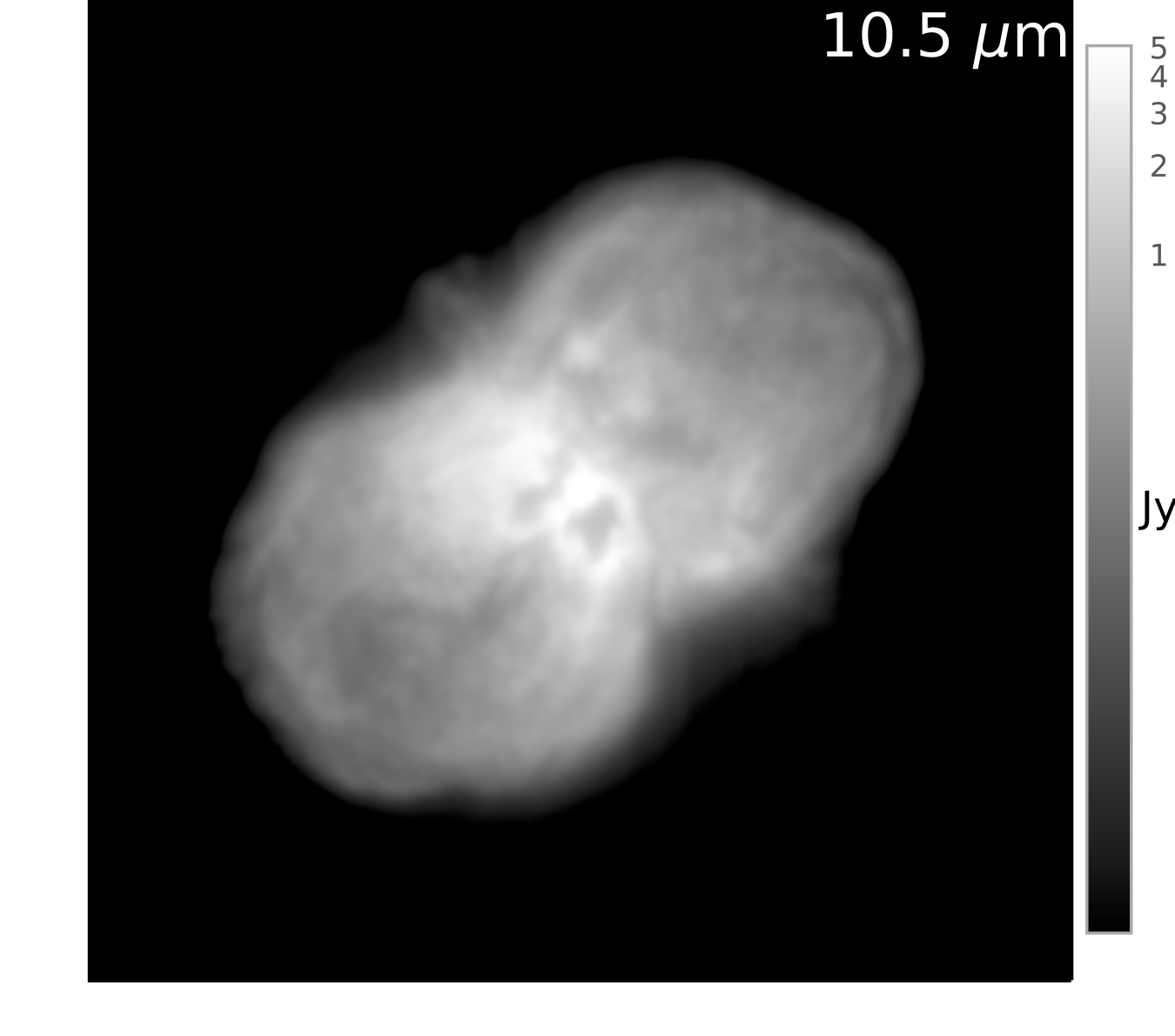}\hspace{0.0\textwidth}
\includegraphics[width=0.34\textwidth]{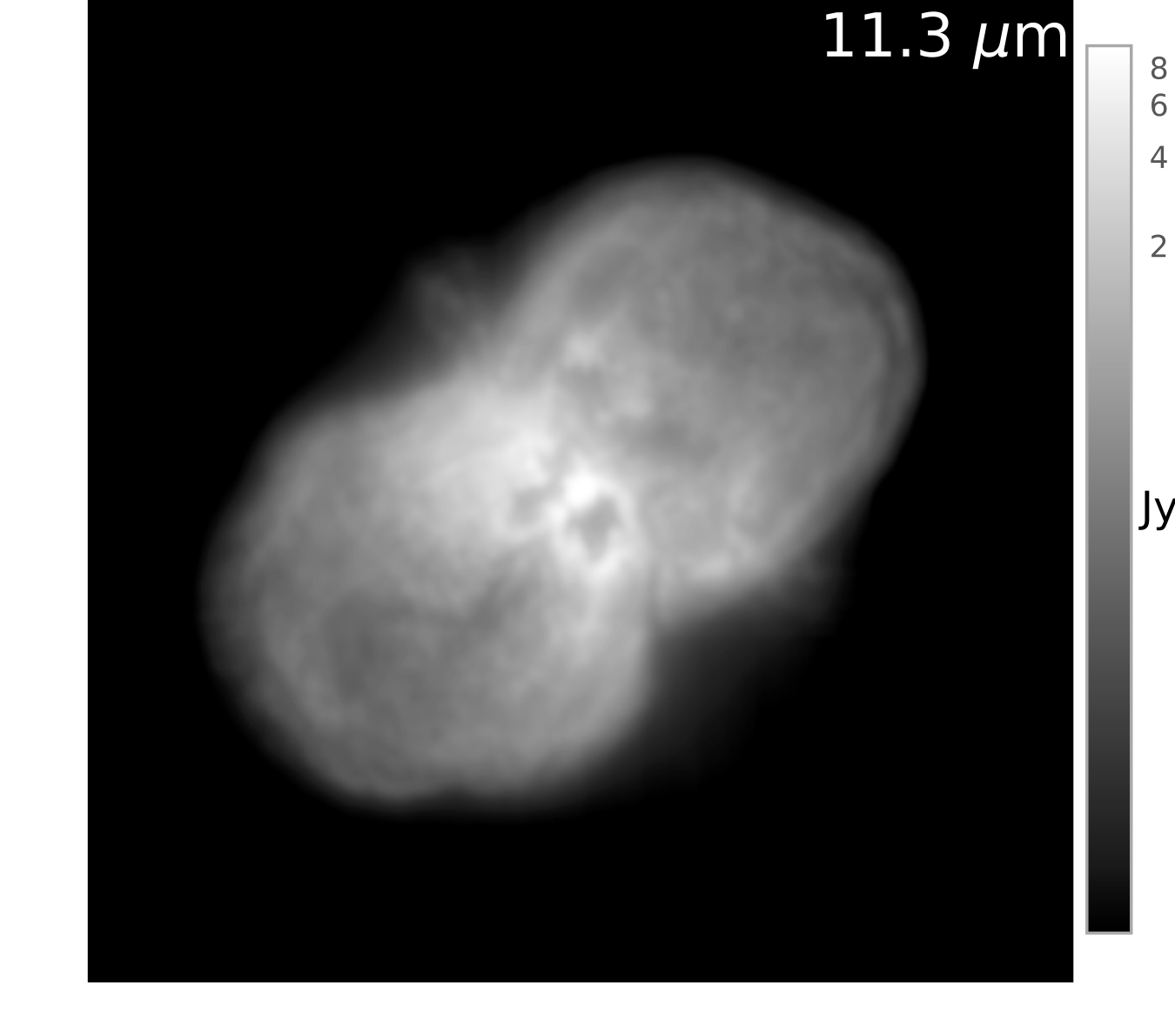}\vspace{0.0\textwidth}
\end{minipage}
\begin{minipage}{0.96\textwidth}
\includegraphics[width=0.34\textwidth]{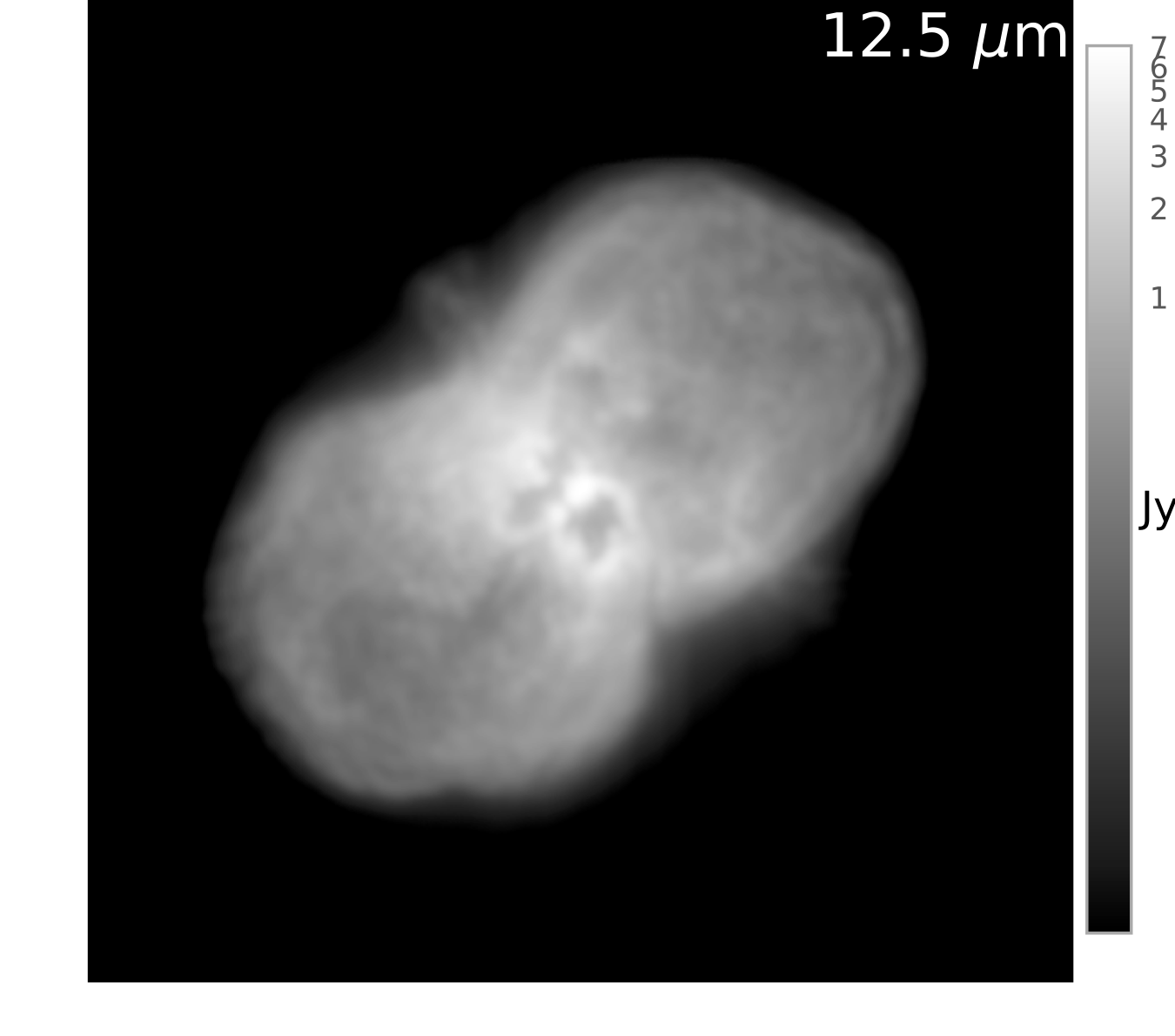}\hspace{0.0025\textwidth}
\includegraphics[width=0.3425\textwidth]{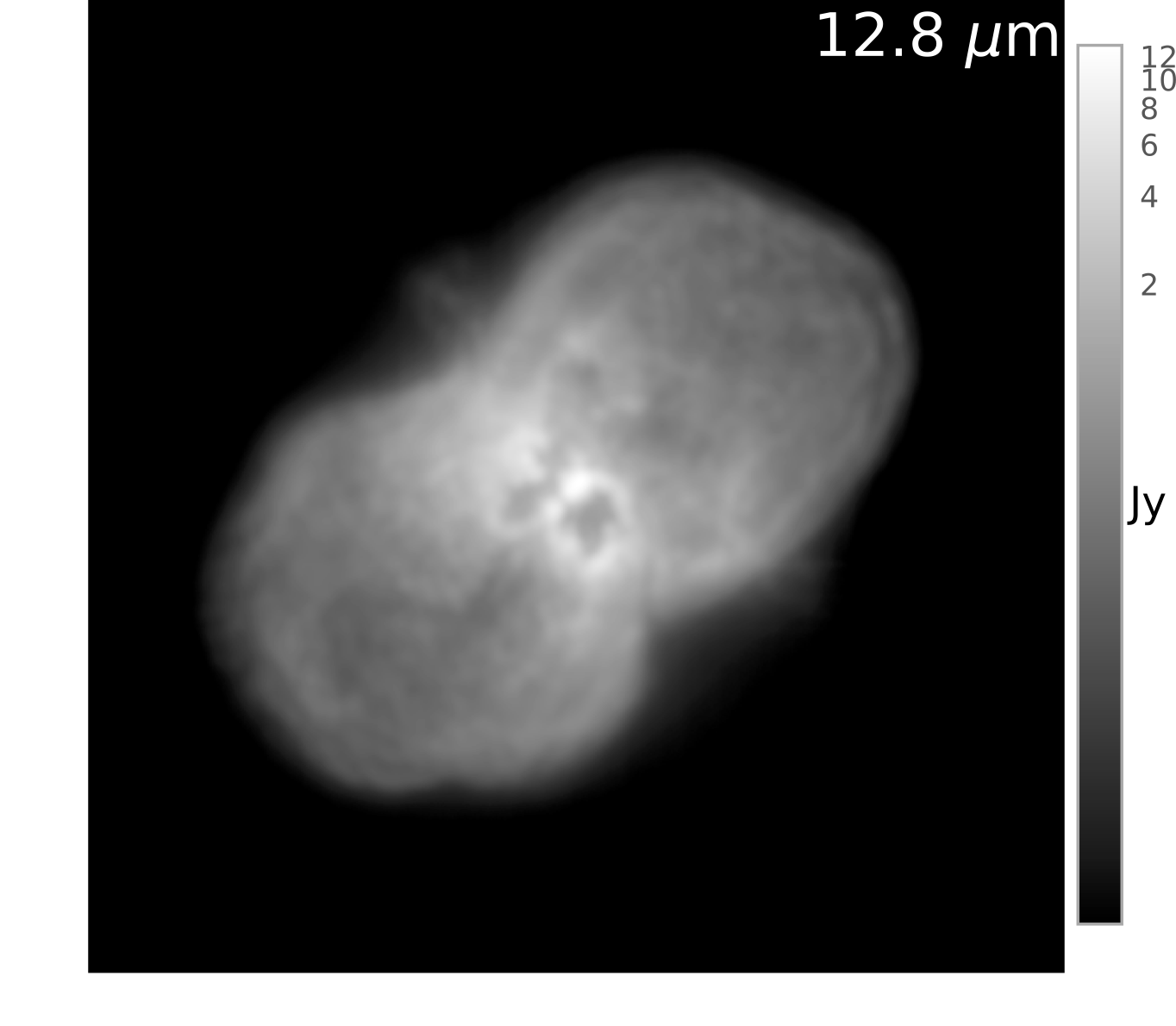}\hspace{-0.0025\textwidth}
\includegraphics[width=0.34\textwidth]{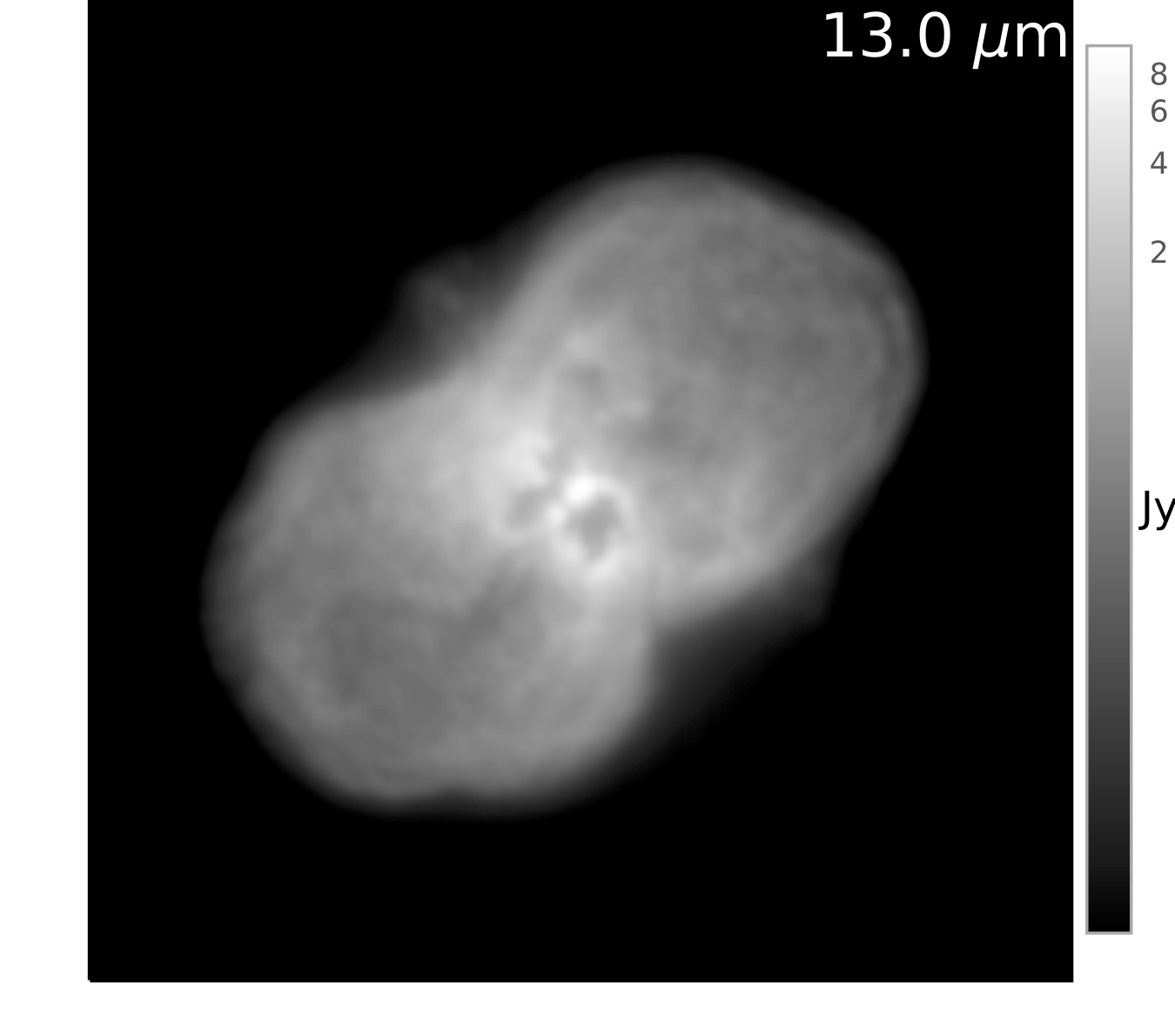}\vspace{0.0\textwidth}
\end{minipage}
\begin{minipage}{0.87114341\textwidth}
\includegraphics[width=0.34\textwidth]{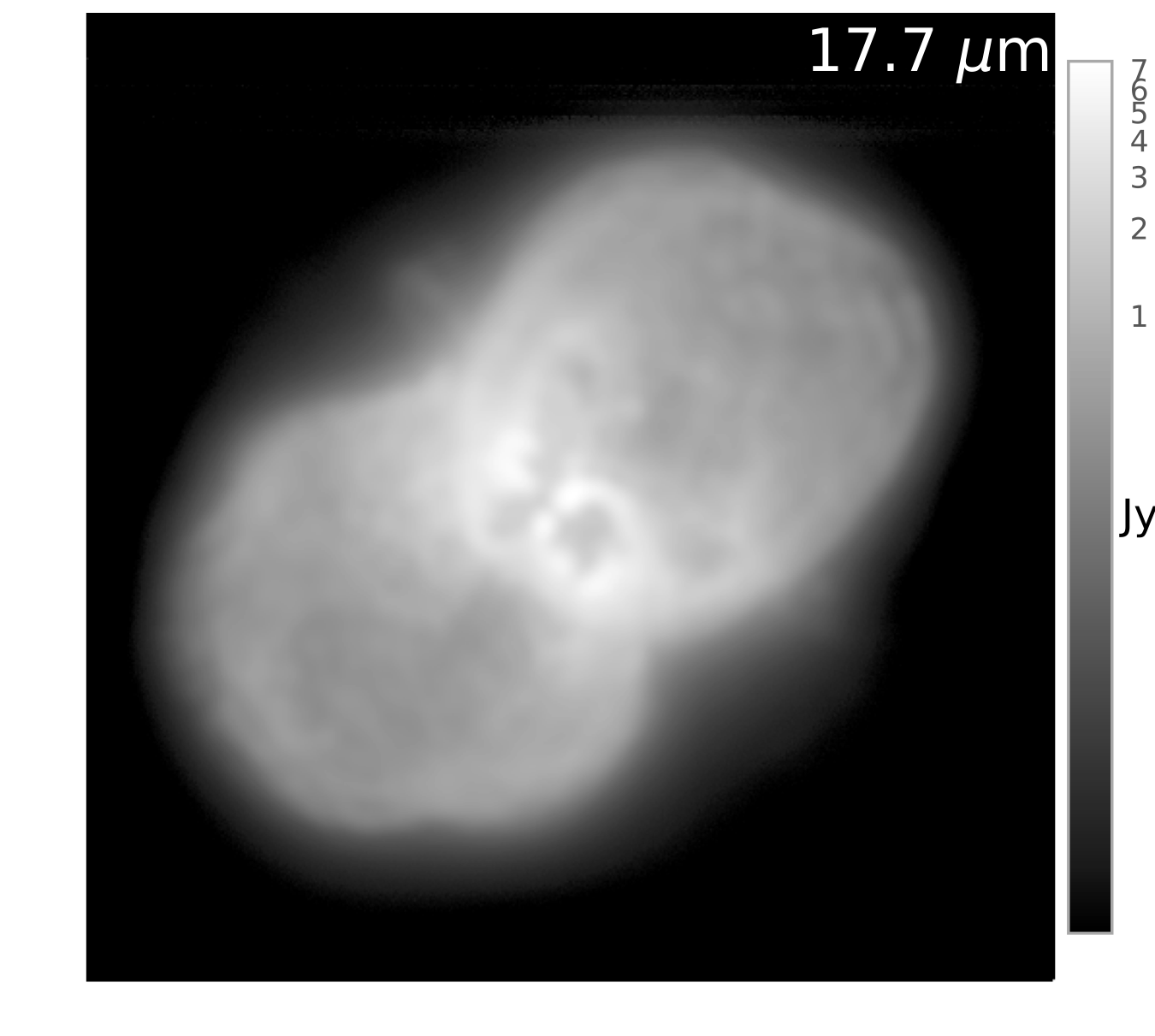}\hspace{0.0\textwidth}
\includegraphics[width=0.34\textwidth]{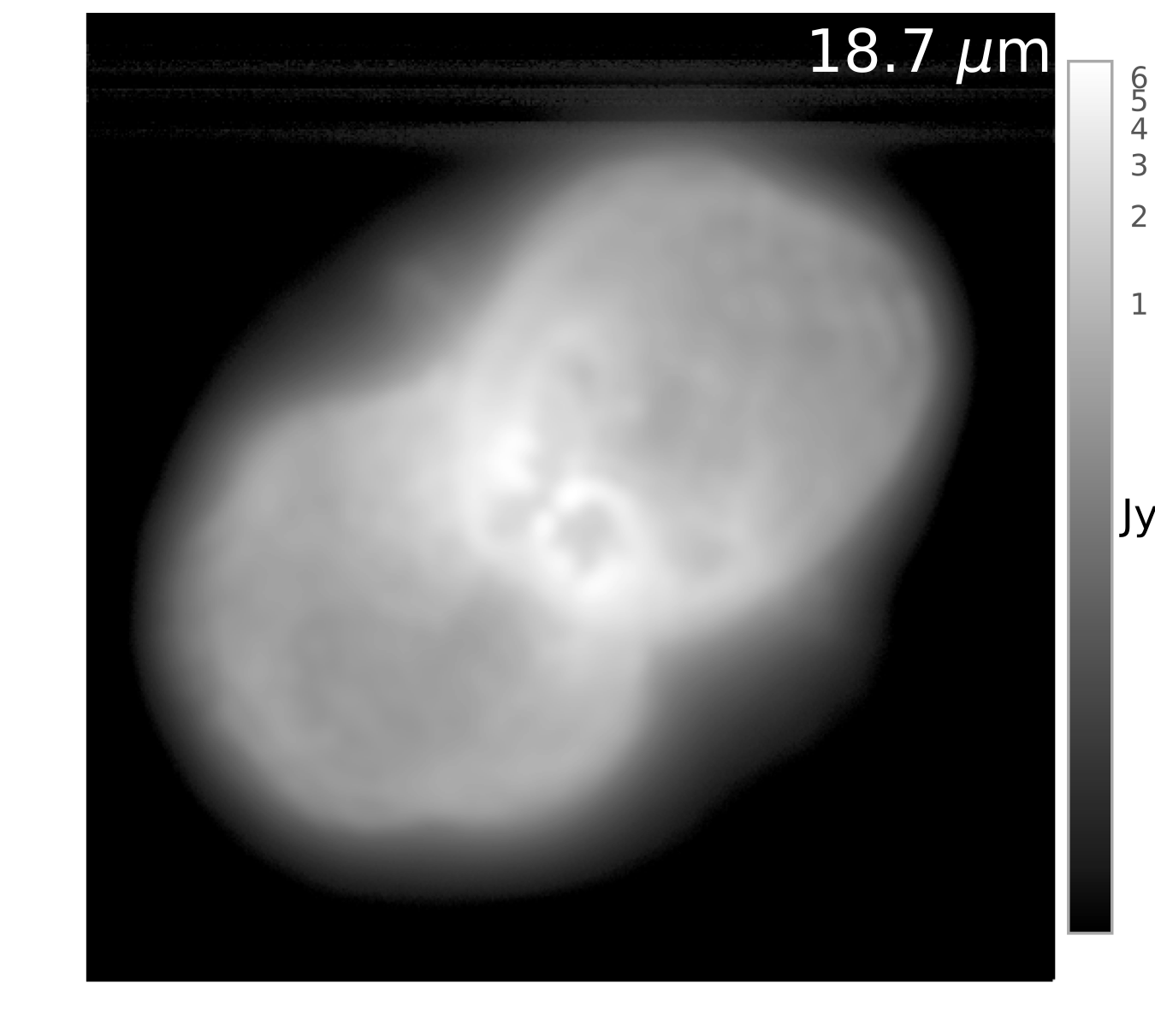}\hspace{0.0\textwidth}
\includegraphics[width=0.34\textwidth]{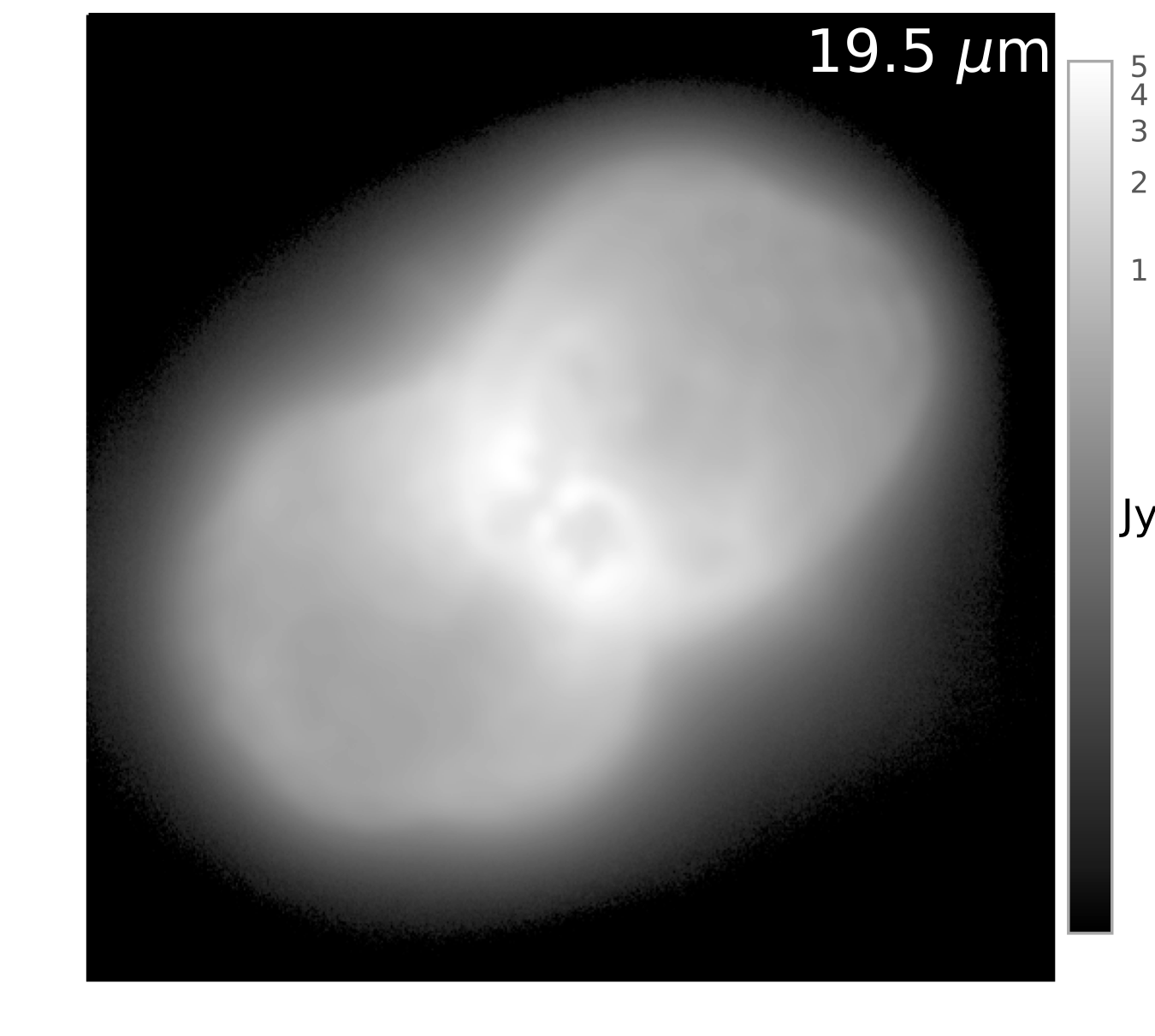}\vspace{0.0\textwidth}
\caption{VISIR 2018 images of $\eta$~Car's Homunculus nebula. The flux density in Jy per detector pixel ($0.045\arcsec$/pixel) is shown on a logarithmic scale. The $N$-band images have a field-of-view of $25\arcsec \times 25\arcsec$, the $Q$-band images (last row) of $22.5\arcsec \times 22.5\arcsec$.} 
\label{figure: VISIR images}
\end{minipage}
\end{figure}

\begin{absolutelynopagebreak}
\section{Journal of VISIR and TIMMI2 observations}
\label{appendix:journal}
\begin{longtable}{lcccccc}
\caption{Journal of VISIR observations.\label{table:journal_visir}} \\
\hline\hline
Date & MJD	&	Target & Filter & Wavelength & Image quality	&	ESO PROG.ID \\ % table heading
   &     &	&	&  ($\mu$m) & (\arcsec) & \\
\hline
\endfirsthead
\hline\hline
Date & MJD	&	Target & Filter & Wavelength & Image quality	&	ESO PROG.ID \\ % table heading
   &     &	&	&  ($\mu$m) & (\arcsec) & \\
\hline
\endhead
2005-01-23	&	53393.05	&	HD9138   &       	PAH2   & 11.25	&	0.39	&         	       	       	074.A-9016(A); PI: Lagage	\\
2005-01-23	&	53393.10	&	HD59294   &      	PAH2   & 11.25	&	0.36&                  	       	       	074.A-9016(A)	\\
2005-01-23	&	53393.18	&	HD30080   &      	NEII   &	12.81	&	0.38	&                  	       	       	074.A-9016(A)	\\
2005-01-23	&	53393.25	&	HD62902   &      	NEII   & 12.81	&	0.38	&                	       	       	074.A-9016(A)	\\	
2005-01-23	&	53393.37	&	eta Car   &      	NEII   & 12.81	&	0.38	&                 	       	       	074.A-9016(A)	\\
2005-01-23	&	53393.37	&	eta Car   &      	PAH2   & 11.25	&	0.37	&                  	       	       	074.A-9016(A)	\\
2018-04-28	&	58237.00	&	eta Car   &	PAH1   &	8.59   &	0.23&          	0101.D-0077(A); PI: Mehner	\\
2018-04-29	&	58237.00	&	eta Car   &	SIV   &	10.49   &0.27&           	0101.D-0077(A)	\\
2018-04-29	&	58237.00	&	eta Car   &	B12.4   &	12.47   &0.31&         	0101.D-0077(A)	\\
2018-04-29	&	58237.01	&	HD89682   &	PAH1   &	8.59 &0.23   &         	0101.D-0077(A)	\\
2018-04-29	&	58237.02	&	HD89682   &	SIV   &	10.49 &0.27   &          	0101.D-0077(A)	\\
2018-04-29	&	58237.02	&	HD89682   &	B12.4   &	12.47   &0.31   &          0101.D-0077(A)	\\	
2018-05-11	&	58249.10	&	eta Car   &	Q1   &	17.65   &$\sim$0.50&                        	0101.D-0077(A)	\\
2018-05-11	&	58249.11	&	eta Car   &	Q2   &	18.72   &$\sim$0.50&                        	0101.D-0077(A)	\\
2018-05-11	&	58249.12	&	eta Car   &	Q3   &	19.50   &$\sim$0.50&                        	0101.D-0077(A)	\\
2018-05-11	&	58249.13	&	HD89682   &   	Q1   &	17.65   &$\sim$0.50&             	       	       	0101.D-0077(A)	\\
2018-05-11	&	58249.13	&	HD89682   &   	Q2   &	18.72   &$\sim$0.50&                        	       	       	0101.D-0077(A)	\\
2018-05-11	&	58249.14	&	HD89682   &   	Q3   &	19.50   &$\sim$0.50&                        	       	       	0101.D-0077(A)	\\
2018-05-21	&	58259.97	&	eta Car   &	J7.9   &		7.78   & 0.22&         	0101.D-0077(A)	\\
2018-05-21	&	58259.98	&	eta Car   &	SIV\_1   &	9.82   & 0.31&        	0101.D-0077(A)	\\
2018-05-21	&	58259.99	&	HD89682   & 	J7.9  &		7.78   &	0.22   &          	0101.D-0077(A)	\\
2018-05-21	&	58260.00	&	HD89682   &	SIV\_1   &	9.82   &	0.31   &        	0101.D-0077(A)	\\
2018-12-10	&	58462.31&	eta Car   &	ARIII   &	8.99   & 0.26&        	0101.D-0077(A)	\\
2018-12-10	&	58462.31	&	eta Car   &	PAH2   &	11.25   &  0.29&        	0101.D-0077(A)	\\
2018-12-10	&	58462.32	&	eta Car   &	NEII   &		12.81   &   0.33&       	0101.D-0077(A)	\\
2018-12-10	&	58462.32	&	eta Car   &	NEII\_2   &	13.04   &   0.33&     	0101.D-0077(A)	\\
2018-12-10	&	58462.33	&	HD89682   &	ARIII   &	8.99   &		0.26           & 	0101.D-0077(A)	\\
2018-12-10	&	58462.34	&	HD89682   &	PAH2   &	11.25   &	0.29             & 	0101.D-0077(A)	\\
2018-12-10	&	58462.34	&	HD89682   &	NEII   &		12.81   &	0.33   &     	0101.D-0077(A)	\\
2018-12-10	&	58462.35	&	HD89682   & 	NEII\_2   &	13.04   & 0.33   &       	0101.D-0077(A)	\\
\hline
%\multicolumn{5}{l}{$^b$ Observations are strongly impacted by ghosts.} \\
\end{longtable}
\end{absolutelynopagebreak}

\begin{longtable}{lccccc}
\caption{Journal of TIMMI2 observations.\label{table:journal_timmi2}} \\
\hline\hline
Date & MJD	&	Target & Filter & Wavelength	&	ESO PROG.ID \\ % table heading
   &  &   &	&  ($\mu$m) &  \\
\hline
\endfirsthead
\hline\hline
Date & MJD	&	Target & Filter & Wavelength &	ESO PROG.ID \\ % table heading
   &  &   &	&  ($\mu$m) &  \\
\hline
\endhead
2003-01-26	&	52665.26	&	HD81797	&  	N10.4	&	10.3 &     	60.A-9126(C); Engineering run \\
2003-01-26	&	52665.26	&	HD81797	&  	N11.9	&	11.6		&     	60.A-9126(C)	\\
2003-01-26	&	52665.26	&	HD81797	&  	N10.4	&	10.3 &	     	60.A-9126(C)	\\
2003-01-26	&	52665.27	&	HD81797	&  	N11.9	&	11.6		&     	60.A-9126(C)	\\
2003-01-26	&	52665.27	&	HD81797	&  	Q1 & 17.8 &     	60.A-9126(C)	\\
%2003-01-26	&	52665.29	&	eta Car &   M & 4.6 &   	60.A-9126(C)	\\
2003-01-26	&	52665.29	&	eta Car &   N7.9	&	7.9	&	     	60.A-9126(C)	\\
2003-01-26	&	52665.30	&	eta Car &    N8.9	&	8.7	&     	60.A-9126(C)	\\
2003-01-26	&	52665.30	&	eta Car &    N10.4	&	10.3 &	      	60.A-9126(C)	\\
2003-01-26	&	52665.3	&	eta Car &    N11.9	&	11.6		&     	60.A-9126(C)	\\
2003-01-26	&	52665.31	&	eta Car &   N12.9	&	12.3	&	    	60.A-9126(C)	\\
2003-01-26	&	52665.32	&	eta Car &    [NeII] & 12.8 &  	60.A-9126(C)\\	
2003-01-26	&	52665.32	&	eta Car &    Q1 & 17.8 &     	60.A-9126(C)	\\
2003-01-26	&	52665.33	&	eta Car &    N9.8	&	9.6	&	      	60.A-9126(C)	\\
2003-01-26	&	52665.33	&	eta Car &    N9.8	&	9.6	&	     	60.A-9126(C)	\\
%2003-01-26	&	52665.34	&	HD81797	&  	M & 4.6 &     	60.A-9126(C)	\\
2003-01-26	&	52665.34	&	HD81797	&  	N7.9	&	7.9	&	      	60.A-9126(C)	\\
2003-01-26	&	52665.35	&	HD81797	&  	N8.9	&	8.7	&	     	60.A-9126(C)	\\
2003-01-26	&	52665.35	&	HD81797	&  	N9.8	&	9.6	&	     	60.A-9126(C)	\\
2003-01-26	&	52665.35	&	HD81797	&  	N10.4	&	10.3 &	     	60.A-9126(C)	\\
2003-01-26	&	52665.35	&	HD81797	& 	N11.9	&	11.6		&     	60.A-9126(C)	\\
2003-01-26	&	52665.35	&	HD81797	&  	N12.9	&	12.3	&	   	60.A-9126(C)	\\
2003-01-26	&	52665.36	&	HD81797	&  	[NeII] & 12.8 &  	60.A-9126(C)	\\
2003-01-26	&	52665.36	&	HD81797	&  	Q1 & 17.8 &    	60.A-9126(C)	\\
%2003-03-19	&	52717.12	&	eta Car &     	N11.9	&	11.6		&     	70.C-0533(A)	; PI: Waters\\
2003-03-19	&	52717.45	&	HD156277	&     	N10.4	&	10.3 &	     	60.A-9126(E); Engineering run	\\
%2003-03-19	&	52717.45	&	HD156277	&     	N11.9	&	11.6		&    	60.A-9126(E)	\\
2003-03-19	&	52717.45	&	HD156277	&     	[NeII] & 12.8 &   	60.A-9126(E)	\\
%2003-05-14	&	52773.05	&	HD123139	&     	N11.9	&	11.6		&     	60.A-9126(A); Engineering run	\\
%2003-05-14	&	52773.06	&	HD123139	&     	N11.9	&	11.6		&     	60.A-9126(A)	\\
2003-05-14	&	52773.07	&	HD123139	&     	N10.4	&	10.3 &	     	60.A-9126(A); Engineering run	\\
2003-05-14	&	52773.07	&	HD123139	&     	N9.8	&	9.6	&	    	60.A-9126(A)	\\
2003-05-14	&	52773.08	&	HD123139	&     	N8.9	&	8.7	&	    	60.A-9126(A)	\\
2003-05-14	&	52773.08	&	HD123139	&     	N7.9	&	7.9	&	   	60.A-9126(A)	\\
2003-05-14	&	52773.08	&	HD123139	&     	N12.9	&	12.3	&	   	60.A-9126(A)	\\
2003-05-14	&	52773.09	&	HD123139	&     	[NeII] & 12.8 &  	60.A-9126(A)	\\
2003-05-14	&	52773.10	&	HD123139	&     	M & 4.6 &    	60.A-9126(A)	\\
%2003-05-14	&	52773.10	&	HD123139	&     	Q1 & 17.8 &     	60.A-9126(A)	\\
2003-05-14	&	52773.12	&	eta Car &      	N7.9	&	7.9	&	     	60.A-9126(I); Engineering run	\\
2003-05-14	&	52773.12	&	eta Car &       	N7.9	&	7.9	&	    	60.A-9126(I)	\\
2003-05-14	&	52773.12	&	eta Car &       	N8.9	&	8.7	&	    	60.A-9126(I)	\\
%2003-05-14	&	52773.13	&	eta Car &       	N9.8	&	9.6	&    	60.A-9126(I)	\\
2003-05-14	&	52773.13	&	eta Car &      	N10.4	&	10.3 &	   	60.A-9126(I)	\\
%2003-05-14	&	52773.13	&	eta Car &      	N11.9	&	11.6		&     	60.A-9126(I)	\\
2003-05-14	&	52773.13	&	eta Car &      	N12.9	&	12.3	&	    	60.A-9126(I)	\\
2003-05-14	&	52773.13	&	eta Car &       	[NeII] & 12.8 &   	60.A-9126(I)	\\
%2003-05-14	&	52773.14	&	eta Car &      	Q1 & 17.8 &     	60.A-9126(I)	\\
2003-05-14	&	52773.14	&	eta Car &      	M & 4.6 &     	60.A-9126(I)	\\
2003-05-14	&	52773.14	&	eta Car &      	N7.9	&	7.9	&	      	60.A-9126(I)	\\
2003-05-14	&	52773.15	&	eta Car &     	N8.9	&	8.7	&	     	60.A-9126(I)	\\
2003-05-14	&	52773.15	&	eta Car &      	N7.9	&	7.9	&	     	60.A-9126(I)	\\
2003-05-14	&	52773.15	&	eta Car &      	N8.9	&	8.7	&	     	60.A-9126(I)	\\
2003-05-14	&	52773.15	&	eta Car &       	N9.8	&	9.6	&	     	60.A-9126(I)	\\
2003-05-14	&	52773.15	&	eta Car &       	N10.4	&	10.3 &	     	60.A-9126(I)	\\
%2003-05-14	&	52773.16	&	eta Car &      	N11.9	&	11.6		&     	60.A-9126(I)	\\
2003-05-14	&	52773.16	&	eta Car &     	N12.9	&	12.3	&	    	60.A-9126(I)	\\
2003-05-14	&	52773.16	&	eta Car &       	[NeII] & 12.8 &  	60.A-9126(I)	\\
%2003-05-14	&	52773.16	&	eta Car &       	Q1 & 17.8 &     	60.A-9126(I)	\\
2003-05-14	&	52773.16	&	eta Car &       	M & 4.6 &     	60.A-9126(I)	\\
2003-05-14	&	52773.96	&	HD108903	&     	N7.9	&	7.9	&	     	69.D-0304(B); PI: Gehrz	\\
2003-05-14	&	52773.97	&	HD108903	&     	N8.9	&	8.7	&	      	69.D-0304(B)	\\
2003-05-14	&	52773.97	&	HD108903	&     	N9.8	&	9.6	&	     	69.D-0304(B)	\\
2003-05-14	&	52773.97	&	HD108903	&     	N10.4	&	10.3 &	     	69.D-0304(B)	\\
2003-05-14	&	52773.97	&	HD108903	&     	N11.9	&	11.6		&     	69.D-0304(B)	\\
%2003-05-14	&	52773.97	&	HD108903	&     	N12.9	&	12.3	&	    	69.D-0304(B)	\\
%2003-05-14	&	52773.97	&	HD108903	&     	[NeII] & 12.8 &  	69.D-0304(B)	\\
2003-05-14	&	52773.97	&	HD108903	&     	Q1 & 17.8 &    	69.D-0304(B)	\\
2003-05-14	&	52773.98	&	HD108903	&     	M & 4.6 &       	69.D-0304(B)	\\
2003-05-14	&	52773.98	&	eta Car &       	N7.9	&	7.9	&	      	69.D-0304(B)	\\
2003-05-14	&	52773.98	&	eta Car &       	N8.9	&	8.7	&	     	69.D-0304(B)	\\
2003-05-14	&	52773.99	&	eta Car &       	N9.8	&	9.6	&	     	69.D-0304(B)	\\
2003-05-14	&	52773.99	&	eta Car &       	N10.4	&	10.3 &	     	69.D-0304(B)	\\
2003-05-14	&	52773.99	&	eta Car &       	N11.9	&	11.6		&    	69.D-0304(B)	\\
2003-05-14	&	52773.99	&	eta Car &      	N11.9	&	11.6		&    	69.D-0304(B)	\\
2003-05-14	&	52774.00	&	eta Car &       	N11.9	&	11.6		&    	69.D-0304(B)	\\
%2003-05-14	&	52774.00	&	eta Car &       N12.9	&	12.3	&	    	69.D-0304(B)	\\
%2003-05-14	&	52774.00	&	eta Car &      	[NeII] & 12.8 &   	69.D-0304(B)	\\
2003-05-15	&	52774.00	&	eta Car &       	Q1 & 17.8 &     	69.D-0304(B)	\\
2003-05-15	&	52774.01	&	eta Car &       	M & 4.6 &      	69.D-0304(B)	\\
2003-05-15	&	52774.01	&	eta Car &       	N8.9	&	8.7	&	     	69.D-0304(B)	\\
2003-05-15	&	52774.01	&	eta Car &       	N11.9	&	11.6		&    	69.D-0304(B)	\\
%2003-05-15	&	52774.01	&	eta Car &       	[NeII] & 12.8 &  	69.D-0304(B)	\\
2003-05-15	&	52774.02	&	eta Car &       	Q1 & 17.8 &     	69.D-0304(B)	\\
2003-05-15	&	52774.94	&	HD108903	&     	N8.9	&	8.7	&	     	69.D-0304(B)	\\
2003-05-15	&	52774.94	&	HD108903	&     	N11.9	&	11.6		&     	69.D-0304(B)	\\
2003-05-15	&	52774.95	&	HD108903	&     	Q1 & 17.8 &     	69.D-0304(B)	\\
2003-05-15	&	52774.95	&	HD108903	&     	M & 4.6 &     	69.D-0304(B)	\\
2003-05-15	&	52774.99	&	eta Car &       	N7.9	&	7.9	&	     	69.D-0304(B)	\\
2003-05-15	&	52774.99	&	eta Car &       	N8.9	&	8.7	&	     	69.D-0304(B)	\\
2003-05-15	&	52775.00	&	eta Car &       	N9.8	&	9.6	&	    	69.D-0304(B)	\\
2003-05-15	&	52775.00	&	eta Car &       	N10.4	&	10.3 &	     	69.D-0304(B)	\\
2003-05-15	&	52775.00	&	eta Car &       	N11.9	&	11.6		&    	69.D-0304(B)	\\
2003-05-16	&	52775.00	&	eta Car &       	N12.9	&	12.3	&	    	69.D-0304(B)	\\
2003-05-16	&	52775.00	&	eta Car &      	[NeII] & 12.8 & 	69.D-0304(B)	\\
2003-05-16	&	52775.01	&	eta Car &       	Q1 & 17.8 &      	69.D-0304(B)	\\
2003-05-16	&	52775.01	&	eta Car &       	M & 4.6 &       	69.D-0304(B)	\\
2003-05-16	&	52775.02	&	HD108903	&     	N7.9	&	7.9	&	      	69.D-0304(B)	\\
2003-05-16	&	52775.02	&	HD108903	&     	N8.9	&	8.7	&	     	69.D-0304(B)	\\
2003-05-16	&	52775.02	&	HD108903	&     	N9.8	&	9.6	&	    	69.D-0304(B)	\\
2003-05-16	&	52775.02	&	HD108903	&     	N10.4	&	10.3 &	     	69.D-0304(B)	\\
2003-05-16	&	52775.02	&	HD108903	&     	N11.9	&	11.6		&     	69.D-0304(B)	\\
2003-05-16	&	52775.02	&	HD108903	&     	N12.9	&	12.3	&	     	69.D-0304(B)	\\
2003-05-16	&	52775.02	&	HD108903	&     	[NeII] & 12.8 &   	69.D-0304(B)	\\
2003-05-16	&	52775.03	&	HD108903	&     	Q1 & 17.8 &      	69.D-0304(B)	\\
2003-05-16	&	52775.03	&	HD108903	&     	M & 4.6 &       	69.D-0304(B)	\\
2003-05-27	&	52786.94	&	HD110458	&     	N7.9	&	7.9	&	     	71.D-0049(A); PI: Zijlstra	\\
2003-05-27	&	52786.94	&	HD110458	&     	N11.9	&	11.6		&    	71.D-0049(A)	\\
2003-05-27	&	52786.95	&	HD110458	&     	N12.9	&	12.3	&	   	71.D-0049(A)	\\
2003-05-27	&	52786.96	&	eta Car &       N7.9	&	7.9	&	      	71.D-0049(A)	\\
2003-05-27	&	52786.97	&	eta Car &       N11.9	&	11.6		&     	71.D-0049(A)	\\
2003-05-27	&	52786.97	&	eta Car &       N12.9	&	12.3	&	    	71.D-0049(A)\\
\hline
%\multicolumn{5}{l}{$^b$ Observations are strongly impacted by ghosts.} \\
\end{longtable}

\end{document}